%% file: main.tex
\documentclass[runningheads]{llncs}

\def\MyTitle{Galaxy Era: Agent-based Simulation of Execution Tickets}
\def\myAuthor{Pascal Stichler}
\def\myInstitute{ephema labs}
\def\myEmail{pascal@ephema.io}
\def\myDate{29.10.2024}

\input{Preamble/packages}

\title{\MyTitle}

\author{\myAuthor}

\institute{%
\myInstitute \\
\email{\myEmail}
}

\begin{document}

%---------- [Cover Page]
\input{CoverPage/Cover}
\thispagestyle{empty}
\cleardoublepage

%--------------- [SHOW TITLE + AUTHORS + INSTITUTE] 
\maketitle 

%--------------- [TABLE OF CONTENT] 
\setcounter{page}{1}
\tableofcontents

%--------------- [ABSTRACT] 
\clearpage
\input{Sections/00-Abstract}

%--------------- [INTRODUCTION & MOTIVATION] 
\clearpage
\input{Sections/01-IntroductionMotivation}

%--------------- [METHODOLOGY] 
\clearpage
\input{Sections/02-Methodology}

%--------------- [MECHANISM OBJECTIVES AND DESIGN SPACE] 
\clearpage
\input{Sections/03-MechanismObjectivesAndDesignSpace}

%--------------- [THEORETICAL ANALYSIS OF SELECTED MECHANISM DESIGNS] 
\clearpage
\input{Sections/04-TheoreticalAnalysisOfSelectedMechanismDesigns}

%--------------- [SIMULATION] 
\clearpage
\input{Sections/05-Simulation}

%--------------- [MITIGATION OF MULTI-BLOCK MEV] 
\clearpage
\input{Sections/06-MitigationOfMultiBlockMEV}

%--------------- [CONCLUSION] 
\clearpage
\input{Sections/07-Conclusion}

%--------------- [REFERENCES] 
\clearpage
\bibliographystyle{apalike}
\setlength{\bibsep}{9pt}

\bibliography{References/references}
\addcontentsline{toc}{section}{\refname}

%--------------- [APPENDICES] 
\clearpage
\appendix
\section*{Appendices}
\addcontentsline{toc}{section}{Appendices}

%--------------- [APPENDIX A: NUMBER OF EXECUTION TICKETS] 
\input{Sections/A-NumberOfExecutionTickets}

\end{document}

%% file: Preamble/packages.tex
\usepackage[margin=0.9in]{geometry}

\usepackage{ textcomp }
\usepackage{bm}

\usepackage[T1]{fontenc}

\usepackage{amssymb,amsfonts,amsmath,amsthm}
\usepackage{graphicx}
\usepackage{xcolor}

\usepackage{float}

\usepackage[shortlabels,inline]{enumitem}

\usepackage{xspace}
\usepackage{ifthen}

\usepackage{comment}

\usepackage{xcolor}
\definecolor{headerColor}{HTML}{aa142d}
\PassOptionsToPackage{hyphens}{url}
\usepackage{url}

\usepackage{hyperref}

\hypersetup{
    pdftitle={\MyTitle},
    pdfauthor={\myAuthor},
    pdfcreator={\myAuthor},
    colorlinks,
    linkcolor={black},
    citecolor={headerColor},
    urlcolor={headerColor},
    breaklinks=true
}

\setlist[itemize]{label=\textbullet}

\usepackage{array} % For enhanced table alignment
\usepackage{tabularx} % For adjustable column widths
\newcolumntype{Y}{>{\centering\arraybackslash}X}
\newcolumntype{M}[1]{>{\centering\arraybackslash}m{#1}}
\newcolumntype{Z}[1]{>{\arraybackslash}m{#1}}

\usepackage{adjustbox}

\usepackage{multirow}

\usepackage{listings}

\definecolor{codegreen}{rgb}{0,0.6,0}
\definecolor{codegray}{rgb}{0.5,0.5,0.5}
\definecolor{codepurple}{rgb}{0.58,0,0.82}
\definecolor{backcolour}{rgb}{0.95,0.95,0.92}

\lstdefinestyle{mystyle}{
    commentstyle=\color{codegreen},
    keywordstyle=\color{green!50!black},
    numberstyle=\tiny\color{codegray},
    stringstyle=\color{headerColor},
    basicstyle=\ttfamily\small,
    breakatwhitespace=false,         
    breaklines=true,                 
    keepspaces=true,                 
    numbers=left,                    
    numbersep=5pt,                  
    showspaces=false,                
    showstringspaces=false,
    showtabs=false,                  
    tabsize=2,
    frame=tb
}

\usepackage{tablefootnote}

\usepackage{parskip}
\usepackage{natbib}
\renewcommand{\refname}{References}
\makeatletter
\renewcommand{\bibsection}{%
   \section*{\refname%
            \@mkboth{\MakeUppercase{\refname}}{\MakeUppercase{\refname}}%
   }
}
\makeatother

\usepackage{tikz}                   % Graphic creation package
\usetikzlibrary{calc}               % TikZ calculation library
\usetikzlibrary{positioning}        % Advanced positioning in TikZ

%% file: CoverPage/Cover.tex
\begin{titlepage}

%======= [ Paper Title ] ======
\tikz[remember picture, overlay] \node[anchor = south, font = \bf\Huge, text width = \linewidth, inner sep = 0pt, align = left, black] (PaperTitle) at ($(current page.center)+(0 cm , 1 cm)$) {\begin{center} \MyTitle \end{center}};

%======= [ Research Report ] ======
\tikz[remember picture, overlay] \node[anchor = north, font = \bf\Large, text width = \linewidth, align = center, black!45] (ResearchReport) at ($(PaperTitle.south)+(0 cm , -1 cm)$) {Research Report};

%======= [ Author + Institute ] ======
\tikz[remember picture, overlay] \node[anchor = south, font = \large, text width = \linewidth, align = center] (AuthorInstitute) at ($(current page.south)+(0 cm , 5 cm)$) {by \myAuthor~(\myInstitute)};

%======= [ Date ] ======
\tikz[remember picture, overlay] \node[anchor = north, font = \large, text width = \linewidth, align = center] (PaperDate) at ($(AuthorInstitute.south)+(0 cm , -0.25 cm)$) {\myDate};

\end{titlepage}

\clearpage

%% file: Sections/00-Abstract.tex
\section{\abstractname}
\label{Abstract}

Execution Tickets present a promising next evolutionary step for enhancing Ethereum's block space allocation mechanism. It separates consensus rewards from execution rewards and sells the execution rights in an effective manner. It aims to foster decentralization among beacon chain validators and enable protocol-level capture of Maximum Extractable Value (MEV). This paper aims to review the theoretical considerations for Execution Tickets, scope holistically the mechanism design space and evaluate potential mechanism design choices based on a structured review and an agent-based simulation.

We develop a theoretical framework identifying three primary objectives of an Execution Ticket mechanism design: decentralization, MEV capture, and Block Producer Incentive Compatibility (BPIC). Further, we propose metrics on how to measure the objectives. For decentralization we propose to use the highest market share, Nakamoto coefficient, and Herfindahl-Hirschman Index, while for MEV capture we propose to measure the MEV share of the protocol from ticket holder rewards. Further, the three price characteristics of price predictability, smoothness and accuracy are identified as desired attributes.

In the next step, seven key mechanism design parameters are identified and explored: ticket quantity, expiry, refundability, resalability, enhanced lookahead, pricing mechanism, and target ticket amount. We propose and investigate four potential pricing mechanisms: First Price Auction (FPA), Second Price Auction (SPA), EIP-1559-style pricing, and AMM-style pricing. Additionally, we propose six concrete mechanism designs based on this parameter design space.

To evaluate the parameters and configurations we implemented an agent-based simulation and based on over 300 simulation runs several findings are concluded. Results indicate that while none of the mechanisms scores particularly well on decentralization, enabling a secondary market reduces centralization by allowing specialized ticket holders to purchase tickets just-in-time. Regarding MEV capture, auction formats and AMM-style pricing performed well, whereas EIP-1559-style pricing captures less MEV. Auction formats with longer lookahead periods demonstrated favorable price predictability and smoothness while scoring slightly less favorable on price accuracy.

Based on this second-price auction format seems most promising as it achieves high MEV capture, adheres to Dominant-Strategy Incentive Compatibility (DSIC) and exhibits favorable price characteristics. Non-expiring tickets score better as they avoid impairing MEV capture due to discounted valuations from expiry risk. Refundability was found to have limited impact on market dynamics and adds complexity; thus, non-refundable tickets are suggested. Embracing a secondary market seems favorable, as it enhances decentralization and increases overall MEV capture. Nevertheless, in line with \citep{bahrani_centralization_2024} we observe that the decentralization of the builder market highly depends on the MEV extraction capabilities of the top builders.

Overall, this study provides a theoretical framework on the mechanism design space for Execution Tickets as well as a practical implementation of an agent-based simulation to test mechanism design choices. Further, it provides an exploratory evaluation of Execution Ticket mechanism designs, offering insights into optimal configurations that balance MEV capture, decentralization, and operational efficiency in Ethereum's block space allocation.

\clearpage

%% file: Sections/01-IntroductionMotivation.tex
\section{Introduction \& Motivation}
\label{sec:IntroductionMotivation}

\subsection{Background}
\label{subsec:Background}

With the London-hard fork on August 5, 2021 a major change was made to Ethereum's transaction fee
mechanism to make transaction fees more predictable and less volatile. Even though it has largely
achieved its purposes \citep{liu_empirical_2022}, some problems of flawed incentives and rewards remain.
Hence, Execution Tickets (ETs), formerly known as Attester-Proposer separation (APS), are currently
discussed as a next step in the evolution of Ethereum's protocol mechanisms, particularly regarding the
transaction fee mechanism (TFM) \citep{neuder_execution_2023}. The first proof-of-stake Ethereum mechanism
envisioned staking as the remuneration for validators \citep{buterin_proof_2022}. With the rise of maximum
extractable value (MEV) validators additionally were able to generate significant remuneration from
MEV gains \citep{daian_flash_2020}. MEV typically results from arbitrage by exploring price discrepancies
between decentralized exchanges (DEX) and centralized exchanges (CEX) and is typically defined as
``excess profit that a validator can extract by adjusting execution of user transactions." \citep{daian_flash_2020,kulkarni_towards_2023}. Generally, MEV is classified into toxic and non-toxic MEV \citep{barczentewicz_mev_2023}. Toxic MEV includes frontrunning and sandwiching transactions, which directly impairs the user.
Non-toxic MEV typically includes backrunning transactions, which however as \citep{barczentewicz_blockchain_2023} argues can also negatively impact users.

Adjusting the transaction ordering to generate MEV is a complex process involving constantly simulating
potential blocks and reducing latency. This leads to centralizing forces as it requires dedicated
investments in the necessary infrastructure \citep{daian_flash_2020}. As a countermeasure, proposer-builder
separation (PBS) was introduced \citep{buterin_endgame_2021}. This separates the block proposer/validator from the
block builder constructing the block. Block construction became a centralized market \citep{oz_who_2024}
and block proposers earned significant additional rewards from auctioning off the block building right to
block builders. This led to proposers earning more rewards than originally intended. Further, it leads to
timing games by validators to stretch the period to receive bids and thereby earn more MEV rewards
\citep{schwarz-schilling_time_2023} and lastly to an increasing centralization of block builders \citep{yang_decentralization_2024}.

Execution Tickets aim to tackle the shortcomings of the current mechanism by separating slots into
beacon and execution rounds with different attesting committees. The beacon block would include the
\href{https://github.com/ethereum/consensus-specs/blob/bf09b9a7c4a7b311e86823235815daf31b117574/specs/capella/beacon-chain.md#beaconblockbody}{beacon block} of today, however without the \href{https://github.com/ethereum/consensus-specs/blob/bf09b9a7c4a7b311e86823235815daf31b117574/specs/bellatrix/beacon-chain.md#executionpayload}{ExecutionPayLoad}. The execution block will consist of the
ExecutionPayLoad including the set of transactions that get included on-chain \citep{neuder_execution_2023}. This is
primarily done to (a) foster decentralization among validators and (b) to capture MEV at the protocol
level. Therefore, MEV rewards are detangled from the beacon chain validator payoffs. The validator set
will still be entitled to beacon block production. However, for the execution round, slots to propose blocks
have to explicitly be purchased from the protocol, e.g. by buying execution tickets. To ensure incentive
alignments and no misbehavior the purchasing mechanism needs to be carefully designed.

\subsection{Structure of the Paper}
\label{subsec:StructureOfThePaper}

We explore this by using a top-down approach of first analyzing the overall mechanism requirements,
then the design mechanism space including specific ticket attributes, investigating pricing and simulating
different configurations. In more detail, we will follow the following structure. In chapter \ref{sec:Methodology} the theoretical
work on the topic will be reviewed and evaluated. In chapter \ref{sec:MechanismObjectivesAndDesignSpace} the objectives of the Execution Ticket
mechanism will be discussed and metrics to measure the objectives will be proposed. Further, the design
space of the mechanism will be laid out. In chapter \ref{sec:TheoreticalAnalysisOfSelectedMechanismDesigns} selected mechanism designs will be proposed and
evaluated on a theoretical basis. In chapter \ref{sec:Simulation} an agent-based simulation framework will be introduced and
used to evaluate the proposed mechanisms as well as the individual mechanism design parameter choices.
chapter \ref{sec:MitigationOfMultiBlockMEV} discusses the topic of multi-block MEV as a particularly crucial risk and potential mitigation
strategies. chapter \ref{sec:Conclusion} concludes and summarizes the work.

\subsection{Introduction to Execution Tickets}
\label{subsec:IntroductionToExecutionTickets}

To “firewall off” the decentralized validator set from centralizing forces the current process of electing a
single block proposer from the validator set to produce the beacon block and execution payload is split up
\citep{neuder_execution_2023}. Due to MEV, currently large incentives exist to outsource the execution payload
construction to an external market of searchers and builders. This creates centralizing pressures on the
validators, such as vertical integration, colocation and pooling. To protect the beacon chain validators
from these effects the execution tickets separate the execution payload. Therefore, external entities can
purchase execution tickets from the protocol. Tickets are initially like lottery tickets and not allocated to
specific slots. Every epoch an enshrined lottery is run to select the execution payload proposers from the
set of ticket holders and thereby allocating each slot to a specific ticket. The lottery mechanism is
introduced to reduce the risk of multi-slot MEV \citep{drake_session_2023}.

In more detail, \citep{neuder_execution_2023} proposes the following basic flow per slot:
\begin{enumerate}[itemsep=-1mm]
    \item During the beacon round, the randomly selected beacon proposer has authorization to propose a beacon block.
    \item This proposer proposes the beacon block that contains the inclusion list.
    \item The beacon attesters vote on the validity and timeliness of the beacon block.
    \item During the execution round, a randomly selected execution ticket has authorization to propose an execution block.
    \item The owner of the ticket is the execution proposer and proposes an execution block.
    \item The execution attesters vote on the timeliness and validity of the execution block.
\end{enumerate}

This allows the beacon chain validators to decentralize while execution block proposers can further
specialize in pricing future slots, risk management, low-latency connections etc. As \citep{buterin_endgame_2021}
writes: “\textit{Block production is likely to become a specialized market.}”

Besides countering centralization effects among beacon chain validators and capturing MEV at protocol
level, ETs will have further positive effects. Due to the lottery, it will cause reward smoothing among
validators and prevent potential MEV stealing (“rugpooling”) by centralized pooling entities. Further, it
will enable easier pre-confirmations for a better user experience and allow to deprioritize single secret
leader election (SSLE) and potentially ePBS \citep{drake_session_2023}.

\citep{burian_mev_2024} has provided a theoretical framework for modeling execution tickets and ran an initial
analysis. One of the main outcomes shows that when Execution Tickets are priced correctly they can
indeed internalize almost all value generated from proposing execution payloads.

\subsection{Limitations}
\label{subsec:Limitations}

In our research we will not focus on the beacon round attestation and the secondary effects ETs might
have on it. One significant effect on validators will be an expected reduction in staking rewards, as they
will not receive the MEV rewards anymore, which as of today drives roughly 10\% of the block rewards\footnote{From \url{https://mevboost.pics/} last retrieved on 24/10/2024}.

Additionally, we will not focus on the specific details of inclusion lists. It will be briefly discussed but in
the simulation and configurations it will not be a focus of the work. Further, timing games are not
included in the simulation. Additionally, we work with static demand functions of the ticket holders that
do not take into consideration the bids of other ticket holders. In addition, considerations around private
order flow are not modeled in the simulation. Furthermore, the role of relays is left out and we don't
simulate missed blocks and missed block penalties.

Additionally, we will not focus on the implementation details of Execution Tickets, how they can be
integrated into the current Ethereum clients and potential issues that might be with the consensus
mechanism.

Regarding the pricing mechanisms we propose initial versions of how they can be designed, however
leave the verification and formal definition to future research. This includes the more in-depth research of
specific parameters such as adjustment steps for EIP-1559-style pricing and others. We only look at this
from an exploratory perspective.

Further, we exclude a more in-depth analysis around the burning mechanism of the earnings from the
Execution Ticket sales. As outlined in \citep{roughgarden_transaction_2020} burning mechanisms usually impair the
OCA-proofness of mechanisms.

\subsection{Related Work}
\label{subsec:RelatedWork}

\subsubsection{Maximum Extractable Value (MEV)}
\label{subsubsec:MaximumExtractableValueMEV}

MEV (maximum extractable value) has been a major concern since \citep{daian_flash_2020} analyzed it. It had
been originally introduced as miner extractable value and described as miners \textit{ordering optimization (OO)
fees}, where miners deviate from the default ordering by price and nonce to extract extra value \citep{daian_flash_2020}. It is relevant to note that also in traditional finance similar forms of arbitrage, such as
frontrunning exist \citep{bernhardt_front-running_2008}, however have been tightly regulated in a form that is
difficult to emulate in DeFi.

\citep{park_unraveling_2024} classifies MEV into three different types with five and two sub-categories. General
arbitrage, sandwich attacks and liquidations. Arbitrage can be further subdivided into Simple Loop
Arbitrage, Burn \& Mint Mechanism Arbitrage, Set Token Arbitrage, Multi Address Arbitrage and NFT
Arbitrage. Sandwich attacks can be sub-classified into Single DEX sandwich and Cross-DEX Sandwich.
Measuring MEV has been historically done by monitoring addresses that are suspected of MEV
generating behavior, especially using high value gas prices as an indicator \citep{daian_flash_2020}. A more
progressive approach has been proposed by \citep{park_unraveling_2024} to use graphic neural networks for
detecting MEV without the need to pre-register new services. \citep{torres_rolling_2024} have recently shown
that also on L2 rollups MEV becomes more prevalent, however the vast majority of MEV (about 10x)
still happens on the L1 Ethereum network.

\subsubsection{Multi-Block Maximum Extractable Value (MMEV)}

One concern raised more recently is around MEV that can be extracted when a party controls more than
one consecutive slot. It was first introduced by \citep{babel_clockwork_2021} as \textit{k-MEV} and further elaborated by
\citep{mackinga_twap_2022}. The concept of \textit{k-MEV} describes the opportunity by a block proposer to generate
MEV by proposing $k$ blocks. \citep{babel_clockwork_2021} further extend the concept to WMEV (weighted MEV),
which weights the MEV by the probability of a block proposer controlling multiple slots. It is defined as:
\[ \textit{WMEV} ( P , s ) = \sum_{k = 1}^{\infty} p_k \cdot k - \textit{MEV} ( P , s ) \]

It has been shown that controlling more than one consecutive slot can lead to higher rewards than
controlling two randomly assigned slots. \citep{jensen_multi-block_2023} observe that in the time period 15th of
September 2022 (the 'Merge') to the 31st of January 2023 the number of consecutive slots by one builder
are significantly higher than a random monte carlo simulation would indicate. Further, \citep{jensen_multi-block_2023} empirically observe indicatively that builders are willing to pay above average for consecutive
slots. As they note, the data is still exploratory and not definitely conclusive, but it shows that there are
already shortly after the Merge indicators for multi-block MEV. A more recent study by \citep{stichler_does_2024}
concludes that in more recent times this trend has reverted and less than expected multi-slot sequences
occur. Nevertheless, they observe that the value for longer slot sequences increases. The reason why this
is happening is still being actively discussed.

\citep{jensen_multi-block_2023} define \textit{collusive MMEV} as a situation where the block proposer and builder collude to
extract MMEV, while \textit{non-collusive MMEV} exists when a builder secures multiple consecutive slots by
winning the MEV-boost auctions. \citep{jensen_multi-block_2023} assume that \textit{collusive MMEV} is constrained as
social norms and a reputational risk for staking pool operators prevent this. Further \citep{jensen_multi-block_2023}
define two MMEV strategies: (i) Discrete strategies, in which builders seek to reach a specific threshold
value to provoke a certain scenario, e.g. liquidation. (ii) Continuous strategies in which the builder seeks
to structure transaction flow throughout the duration by ordering transactions. Continuous strategies are
likely the favorable option, as they may fail without significant cost to the builder.

A specific form of multi-block MEV in terms of attacking lending protocols is outlined by \citep{mackinga_twap_2022}. In this case an attacker can use the control over multiple blocks to manipulate time-weighted
average price (TWAP) oracles. \citep{mackinga_twap_2022} therefore classify the necessary capital
requirements into “attack capital” and “manipulation capital”. The strategies include:
\begin{enumerate}[label = (\alph*)]
    \item \textbf{\textit{Undercollateralized loan attack:}} In this case, a malicious actor uses capital to artificially
    inflate the price of an asset A by buying it from an AMM. In the next step, a loan is
    borrowed from a lending protocol that uses the AMM to inform its own collateralization
    ratio. Now the attacker can use further capital as collateral on the lending protocol and
    borrow a loan asset B. In the next step, the attacker sells all assets and due to the inflated
    price of A, the loan is undercollateralized. An example of this, even though not strictly to
    be classified as MMEV, was the attack on Inverse Finance DAO's anchor lending
    protocol, that caused a loss of 15.6mn USD\footnote{\url{https://rekt.news/inverse-finance-rekt/} retrieved on 25/02/2024}.
    \item \textbf{\textit{Liquidation Attack:}} In this setting a bad actor artificially drives down the price of asset A
    in block $N$ to then liquidate collateralized loans on Asset A in block $N + 1$.
\end{enumerate}

A major risk for multi-block MEV strategies is the possibility of chain reorganizations (“reorgs”).
Especially, for MMEV strategies that involve investing capital in block $N$ to profit in block $N + 1$ the risk
that block $N + 1$ is reorged is notable. This reorg might either be “honest” or “malicious”. In the case of a
reorg the “manipulation capital” might be lost or traded off at a loss. This could lead to an equilibrium in
which the MMEV opportunity is limited as an upper bound by the costs to do a malicious reorg of the
chain. To better understand this, it is relevant to understand the cost of a reorg. \citep{neuder_execution_2023} and
\citep{schwarz-schilling_three_2022} have outlined different attack vectors and the respective costs. Based on
this it can be concluded that no techniques are known for ex post attacks for a non–majority holding
adversary. Given this, the risk of a proposer controlling multiple slots getting the last slot maliciously
reorged is low, as this would need to be done ex post.

In the context of Execution Tickets multi-block MEV is of particular importance as the mechanism
changes from just-in time (JIT) auctions of the current \href{https://github.com/flashbots/mev-boost}{MEV-Boost} implementation to a mechanism with a
potentially longer lookahead period for builders. For example, \citep{neuder_execution_2023} and \citep{burian_future_2024} cite this
as one of the major concerns. It is to note that already in the current market structure block validators
theoretically have the certainty that they periodically receive multi-slot sequences, e.g. as \citep{stichler_does_2024}
have found, the longest observed sequence since the introduction of proof-of-stake with the same
validator and builder so far was 11 slots on March 4th, 2024 by Lido \& BeaverBuild. Especially
validators with local block-building could already run multi-slot MEV strategies. However, \citep{stichler_does_2024} has looked into data since 15th of September 2022 (the 'Merge') and has not observed patterns of
multi-slot MEV. Execution Tickets, particularly when a secondary market exists, allow for deliberately
purchasing multi-slot sequences at a certain point in time which opens up the playground for more
variations of multi-slot MEV strategies.

\subsubsection{Builder Market}
\label{subsubsec:BuilderMarket}

\citep{yang_decentralization_2024} have shown that there are significant centralizing forces in the builder market. One of
the main centralizing effects is the access to order flow. Similar to traditional finance, payment for order
flow has become a significant share of the market as builders compete to get the most proprietary order
flow. \citep{yang_decentralization_2024} outline a spin-wheel effect that builders with large market shares have easier
access to further private order flow which in turn enables them to have an even higher market share. In
order to gain market share small builders subsidize their auctions and bid higher than their extracted 
values, which leads to only 79\% of auctions being efficient in the classical economical sense.

They divide builders into top (1-5), middle (6-15) and tail (16-25) builders based on market share in the
time period April to August 2023 and show that the block building capabilities (ability to extract) MEV
differs intergroup however not so much intragroup.

\clearpage
\subsection{Pricing model of Execution Tickets}
\label{subsec:PricingModelOfExecutionTickets}

To understand the motivations and actions of market participants we deem it feasible to outline how
Execution Ticket buyers will value the tickets. \citep{burian_execution_2024} has given a first overview of how execution
tickets can be valued and \citep{schlegel_pricing_2024} has shown that execution rewards are mean reverting. Further,
we define how execution tickets can be specified and adapt traditional perpetual bond pricing to the
context of ETs.

\subsubsection{Theoretical background}
\label{subsubsec:TheoreticalBackground}

To understand the incentives around Execution Tickets, we outline potential pricing models. Existing
pricing models from option pricing can provide guidance. Therefore, we need to distinguish between
unallocated and allocated ETs. When tickets have no expiration date, unallocated execution tickets have
similar attributes to perpetual bonds in financial markets. Allocated execution tickets share many
similarities with futures, as they are eligible to be a block producer at some point in time.

\paragraph{Perpetual Pricing}
\label{par:PerpetualPricing}

Perpetual bonds have been a financial instrument for a long time, with the first issued bond dating back to
the 15th of May 1648 by the Dutch water board of Lekdijk Bovendams\footnote{Wigglesworth, Robin (2024-02-12). \href{https://www.ft.com/content/9a13d322-bab7-4463-99c6-972cfe0eb4ec}{"The world's oldest living bond"}. Financial Times. Retrieved 2024-02-12.}. They grant the debtor an
indefinite payment of a defined coupon. Perpetuals are typically priced as $ P_{\textit{Perp}} = I / d$ where 
$I$ is a periodic coupon on a bond and $d$ is the discount rate \citep{black_valuing_1976}. 

In the context of ETs, the ET holder of one ticket has the expected payout per slot of $ \mu_R = \dfrac{r}{n} $ at each 
allocation period, where $r$ represents the revenue from validating the block and $n$ the number of
outstanding tickets. Thus, it can be thought of as a perpetual payout at each allocation period. However, in
the context of ETs, a cost to carry per allocation period $c$ needs to be included. By buying the ET, the
validator agrees to be able to validate the execution block and therefore needs to maintain the
infrastructure to validate blocks. Therefore, the value of an unallocated ET can be defined as:
\[ E(V_{\textit{Ticket}}) = \dfrac{(\mu_r - c)}{dn + 1} \]

The proof can be found in \citep{burian_execution_2024}. The discount rate $d$ depends on the cost of capital of the
specific execution ticket holders. It may be approximated further by the historical returns in similar asset
classes.

\paragraph{Future Pricing}
\label{par:FuturePricing}

The moment an ET is allocated, it changes its properties. The validator can be sure to receive a payout $q$
at (or before) its allocated slot. Normally, futures are priced as the spot price plus the cost to carry. In the
context of ETs, the cost to carry can be neglected assuming reasonably short lookahead periods.
Therefore, the price of the future is close to the payout $q$, which can be derived from the value of the
expected MEV. However, as shown by \citep{schwarz-schilling_time_2023} the majority of bids per block
only arrive very short-notice starting at around 2.5 seconds before the slot boundary and increase over
time. This indicates that the expected MEV can only be estimated at very short notice, which has led to
more sophisticated timing games \citep{schwarz-schilling_time_2023}. From this we can derive that the value
of allocated tickets can only be approximated by the expected MEV with a few slots look-ahead. To
verify this historical MEV-Boost payment data is analyzed. \citep{stichler_does_2024} further concluded that there is
a moderate correlation between consecutive slot MEV-Boost payments spanning no more than up to three
slots. The distribution of MEV-Boost payments is heavily skewed and driven by outliers. Based on this, a
rank-based correlation shows higher values than the Pearson coefficient. It can be concluded that even
allocated tickets will generally be priced by the average expected MEV until at least $N - 2$.

This might be different in volatile times of generally high MEV and given the skewed distribution of
MEV also sets a lower baseline for low-MEV times. It is to note that the analysis of \citep{stichler_does_2024} is
limited by using historical data for the prediction of prices. There might be other strategies to predict slots
with high MEV earlier such as monitoring the mempool, volatility and other correlated variables.

%% file: Sections/02-Methodology.tex
\section{Methodology}
\label{sec:Methodology}

Based on the previous literature review, the approach to validate potential mechanism designs for
Execution Tickets is twofold.

In the first step, we conduct a theoretical analysis of the objectives that the mechanism aims to optimize
and propose several metrics to measure the achievement of these objectives. The objectives are divided
into primary and secondary categories. Next, we outline the design space of possible mechanism
attributes and their potential values. This includes properties of the tickets as well as potential pricing and
allocation mechanisms (e.g., auction-based formats vs. quoted-price formats). For each attribute, we
provide a review and reasoning on the impact it might have on the mechanism. Based on this analysis,
potential concrete mechanism designs are proposed and evaluated using a theoretical framework.

In the second step, these findings are tested with an agent-based simulation. Simulations (e.g. \href{https://ethereum.github.io/abm1559/notebooks/stationary1559.html}{EIP-1559
simulation}) have proven to be a suitable tool to estimate the impact of potential mechanism design
choices. The simulation emulates the allocation, trading, and redemption processes of Execution Tickets.
The scope of the simulation is to run the previously designed configurations and compare them based on
the objectives. Furthermore, conclusions can be drawn from the simulation about each parameter to
determine favorable choices. The simulation is implemented in Python using existing industry standard
frameworks (\href{https://github.com/BenSchZA/radCAD}{radCAD}). It is developed as a Jupyter Notebook and hosted on Google Colab.

%% file: Sections/03-MechanismObjectivesAndDesignSpace.tex
\section{Mechanism Objectives and Design Space}
\label{sec:MechanismObjectivesAndDesignSpace}

To outline the possible design mechanisms we first outline the solution space of different configurations
before evaluating them individually. In \autoref{tab:SummaryOfImportantObjectives} we summarize the desired attributes outlined in chapter \ref{subsec:MechanismObjectives}
to \ref{subsec:PricingBehavior}.

\begin{table}[H]
\centering
\renewcommand{\arraystretch}{1.5} % Row height adjustment
\setlength{\arrayrulewidth}{0.15mm} % Thick lines
\setlength{\tabcolsep}{0.5em}

\adjustbox{max width = \linewidth}{
\begin{tabular}{|M{4cm}|Z{6cm}|Z{6cm}|}
    \hline
     & \textbf{Objectives} & \textbf{Measurement metrics}
    \\ \hline

    \textbf{Optimization Parameters}
    & 
    \begin{itemize}[left=0pt, topsep=5pt]
        \item Decentralization
        \item MEV Capture
        \item BIPC
    \end{itemize}
    &
    \begin{itemize}[left=0pt, topsep=5pt]
        \item Market share, Nakamoto-coefficient \& Herfindahl-Hirschman Index
        \item MEV-Share Protocol
    \end{itemize}
    \\ \hline

    \textbf{Operational Process} 
    & Minimal communication, ease of ticket holding, effective penalties
    & Qualitative assessment
    \\ \hline

    \textbf{Pricing Behavior}
    & 
    \begin{itemize}[left=0pt, topsep=5pt]
        \item Price Predictability
        \item Price Smoothness
        \item Price Accuracy
    \end{itemize}
    &
    \begin{itemize}[left=0pt, topsep=5pt]
        \item GK Measure
        \item $V(\Delta p)$
        \item MEV-Share Protocol
    \end{itemize}
    \\ \hline

\end{tabular}
}

\vspace*{2mm}
\caption{Summary of important objectives}
\label{tab:SummaryOfImportantObjectives}

\end{table}

In \autoref{tab:OutlineOfPossibleExecutionTicketConfigurations} we outline the design space of possible configurations of Execution Tickets described in chapter \ref{subsec:TicketParametersAttributes} to \ref{subsec:TargetAmountOfTickets}.

\begin{table}[H]
\centering
\renewcommand{\arraystretch}{1.5} % Row height adjustment
\setlength{\arrayrulewidth}{0.15mm} % Thick lines
\setlength{\tabcolsep}{0.5em}

\adjustbox{max width = \linewidth}{
\begin{tabular}{|Z{6cm}|Z{6cm}|}

    \hline
    \textbf{Ticket Attributes} & \textbf{Configurations} \\ \hline
    Amount of tickets & Variable / Fixed \\ \hline
    Expiring tickets & Yes / No \\ \hline
    Refundability & Yes / No (unallocated \& allocated) \\ \hline
    Resalability & Yes / No (unallocated \& allocated) \\ \hline
    Enhanced Lookahead & No / Yes (x epochs) \\ \hline
    \textbf{Possible Pricing Mechanisms} & EIP-1559 style, AMM style, FPA,
    SPA, Dutch Auction \\ \hline
    \textbf{Target Amount} & \# of tickets (for variable / fixed) \\ \hline

\end{tabular}}

\vspace*{2mm}
\caption{Outline of possible execution ticket configurations}
\label{tab:OutlineOfPossibleExecutionTicketConfigurations}

\end{table}

\subsection{Mechanism Objectives}
\label{subsec:MechanismObjectives}

In the first step we will outline what the goal of the proposed mechanism design is and therefore the
objectives it aims to optimize. As a reminder, Execution Tickets are proposed to foster two objectives: (a)
foster decentralization among validators and (b) to capture MEV at the protocol level. Hence the
mechanism shall be designed to maximize decentralization among beacon chain validators and the share
of MEV captured at protocol level. It is to note that multi-round MEV might result from unfavorable
mechanism design in the sense that one validator captures several successive slots and extracts MEV
thereof. By measuring the MEV share captured at protocol level, this is implicitly included.

Decentralization is a key aspect, as it prevents several other unfavorable dynamics. In the context of
execution tickets it can be divided into beacon chain validator and execution chain proposer
decentralization. Generally, decentralization ensures liveness in the sense that not a single actor can
voluntarily or involuntarily halt the chain and impair liveness. Further, it contributes to censorship
resistance \citep{lee_dq_2021}. For all of these reasons beacon chain validator decentralization is paramount.
Execution chain proposer decentralization is less critical under the assumption that beacon chain
validators can force certain transactions into the block (for more details on this see chapter \ref{subsec:InclusionListsFOCILandAUCIL}.). In this
case, execution chain decentralization is mainly relevant to avoid liveness risk and to ensure competitive
bidding for ETs under the constraint that off chain-collusion can be avoided.

Capturing MEV at the protocol level is essential as it removes MEV rewards from beacon chain validator
rewards and most likely burning the rewards is the most neutral way to do so. As a secondary effect, it
increases the welfare of the token holders. Currently, earnings from MEV only benefit the MEV supply
chain as well as block validators. By capturing MEV at protocol level, it benefits all token holders. If the
earnings are burned, which most likely is the most efficient mechanism, it basically functions as deflation.

Further, the mechanism needs to be Block Producer Incentive Compatible (BPIC). The concept of BPIC
is introduced by \citep{bahrani_transaction_2023} and a special case of Myopic Miners Incentive Compatible
(MMIC) introduced below. A transaction fee mechanism (TFM) is considered BPIC, if it expects a block
producer to publish a block that maximizes its private valuation plus the net fees earned. This is essential,
as in the Ethereum mechanism design block validators cannot be forced to propose blocks. Hence, they
need to be properly incentivized to firstly participate in the network and secondly to not propose empty
blocks.

Regarding the pricing mechanism from an economic perspective, it is crucial to evaluate the feasibility of
potential mechanisms against the criteria outlined by \citep{roughgarden_transaction_2021}. These criteria include
ensuring that the mechanism is Myopic Miners Incentive Compatible (MMIC), Dominant-Strategy
Incentive Compatible (DSIC), and resistant to Off-Chain Agreements (OCA-proof).

The criteria defined by \citep{roughgarden_transaction_2021} can be interpreted as the following:
\begin{itemize}
    \item MMIC suggests that it is always irrational for a miner to create fake transactions. This implies
    that under the MMIC framework, the system is designed to discourage miners from manipulating
    transactions for their own benefit.
    \item OCA-proofness stands for a system's resilience against off-chain alternatives that could
    pareto-improve over a canonical on-chain outcome. This means that if a system is OCA-proof,
    there are no off-chain alternatives that would make at least one participant better off without
    making anyone else worse off compared to the original on-chain outcome.
    \item DSIC refers to a situation where if a miner follows the allocation rule $x$, each user is best off
    playing their dominant strategy regardless of the bids of other users. This concept indicates that in
    systems adhering to DSIC, users have a clear best course of action that does not depend on the
    actions of others. DSIC is commonly associated with good UX, as the end users do not need to
    strategize with their actions.
\end{itemize}

\subsubsection{Measuring the Objectives}
\label{subsubsec:MeasuringTheObjectives}

\paragraph{Decentralization}
\label{par:Decentralization}

Measuring decentralization is notoriously difficult. A simple approach is to measure the count of stake
pools or validators, distribution of the token supply across those validators and the percentage of staked
token supply. It might however, not be trivial to identify several nodes belonging to the same staking pool
/ owner. Further approaches proposed in the literature include measuring censorship resistance and
geographical diversity \citep{lee_dq_2021}. Further \citep{lin_measuring_2021} propose to use Gini coefficient,
Shannon entropy, and Nakamoto coefficient over different time periods to measure decentralization.
Another metric proposed by e.g. \citep{heimbach_ethereums_2023} is the Herfindahl-Hirschman Index (HHI) index.

It is calculated as the sum of the squared market shares of the individual participants such as \( H = \sum^{N}_{i=1} a_i^2 \)
with \( a_i = \dfrac{x_i}{\sum^N_{j=1} x_j} \).

In our case we deem it important to measure the decentralization two fold: on the one hand it applies to
the diversity of the beacon chain validators. Fostering decentralization among beacon chain validators is
imminent and extremely important. However, as the simulation focuses on the execution chain, we will
only evaluate the beacon chain decentralization on a theoretical basis and focus the quantitative
assessment of decentralization on the execution chain. To measure decentralization, we propose: (i)
market-share of the largest ticket holder measured on redeemed tickets. (ii) Nakamoto-coefficient
(minimum number of market participants needed to collude to control 51\%) over the simulation. (iii)
Herfindahl-Hirschman Index (HHI) index for the timespan of the complete simulation to also capture
long-tail decentralization.

\paragraph{MEV Capture}
\label{par:MEVCapture}

As the secondary important objective of ETs is to capture MEV at protocol level, it needs to be carefully
evaluated how this shall be measured. The primary goal of ETs is not to maximize MEV rewards, but to
capture inevitable rewards at protocol level. Hence, we propose to measure the share of MEV rewards
captured. Therefore, we are looking for:
\[ \textit{Share}_{\textit{Protocol}} = \dfrac{\textit{Captured MEV Rewards}}{\textit{Total MEV Rewards}} \]

Measuring MEV rewards is hard, as shown by \citep{judmayer_estimating_2023}. However, as \citep{wahrstatter_time_2023} have outlined, the amount of ETH transferred to validators via MEV-Boost can be used as a good
proxy measure for generated MEV rewards. At time of this writing, this information is published by
Flashbots, Ltd and Toni Wahrstätter\footnote{\url{https://mevboost.pics/} (retrieved on 05/07/2024)}. Currently, MEV-Boost has a consistent market share of around 90\%.
The interesting question remains if the remaining 10\% forgo MEV-rewards or if they use different ways
to capture it, e.g. by MEV-aware local block building if a threshold bid value is not met.

To define the captured MEV rewards we can use the proceeds generated by the execution ticket sale.

\subsection{Operational Process}
\label{subsec:OperationalProcess}

From an operational perspective, the efficiency and effectiveness of ticket selling processes are
paramount, especially in environments where accuracy and reliability are non-negotiable. To this end,
several factors have been identified as critical in optimizing these processes, to ensure they are both
robust and streamlined.

At the forefront of these considerations is the principle of minimal communication requirements during
the ticket selling phase. The essence here is to circumvent the complexities associated with sophisticated
price negotiations, which typically involve extensive back-and-forth communication. Such complexities
are not only time consuming but also increase the susceptibility to interruptions and errors, thereby
compromising the process's overall efficiency. Additionally, this might have secondary effects on the
consensus mechanism making it more complex by more communication rounds.

Another critical factor is the ease of holding execution tickets. This aspect underscores the importance of
a system design that facilitates the seamless management and retention of execution tickets by validators.
This enables less sophisticated entities to participate in the market as well.

Further, it will be desirable to design the mechanism in a way that reduces the need for derivative
products. From a mechanism design perspective derivatives such as e.g. “liquid execution tickets” are not
desirable as they might have properties that decrease the stability and might be a sign that not all value
capture has happened at protocol level.

Lastly, the establishment of a mechanism to penalize validators who miss their slots is vital. Such a
mechanism serves as a deterrent against non-compliance and ensures that all participants adhere to the
mechanism design.

\subsection{Pricing Behavior}
\label{subsec:PricingBehavior}

Regarding the pricing behavior certain aspects are favorable. We deem the following from a protocol
perspective most relevant.

\subsubsection{Price Predictability}
\label{subsubsec:PricePredictability}

The price of Execution Tickets shall be predictable for ticket holders in order to be able to meaningfully
participate in the auctions and also do long term planning. Hence, in the scope of this work we focus on
the long term price predictability.

As summarized in \citep{poon_forecasting_2003} volatility can be a measure for price predictability in financial
markets. Classical methods to measure volatility are using squared returns based on daily opening and
closing prices as outlined in \citep{parkinson_extreme_1980}. However, \citep{tan_speculative_2020} propose to use the GK and RS
measures proposed by \citep{parkinson_extreme_1980}, \citep{garman_estimation_1980} and \citep{rogers_estimating_1991}, as they
are five to seven times more efficient than the squared returns measure. It uses the daily opening price,
daily closing price, daily high and daily low price to construct a measurement for volatility. In the case of
Execution Tickets, measuring volatility depends on the pricing mechanism selected. Generally, we deem
the GK measure suitable as well, however the time periods might need to be adjusted from daily to a
suitable interval, depending on how often ETs are sold. In the simulation we adjust it to an per epoch
base. Furthermore, it does not take the drift of the underlying into account, given the short time spans that
are investigated, we deem this acceptable \citep{bennett_measuring_2012}.

\subsubsection{Price Smoothness}
\label{subsubsec:PriceSmoothness}

Even during periods of demand surges or declines and high market volatility prices for Execution Tickets
shall not be fluctuating extremely. This is necessary to ensure even returns for Execution Ticket holders
and hence reduce the risk/reward-function. However, given the professionality of the involved actors this
might be of secondary importance.

To measure the smoothness of the prices we suggest using the variance of the price delta of consecutive
slots $V(\Delta p)$. Thereby large price fluctuations are reflected while continuous changes will result in a low
variance and a drift of the underlying is reflected. For future use cases, more sophisticated measurements
might be investigated like outlined in \citep{froeb_measuring_1994}.

\subsubsection{Price Accuracy}
\label{subsubsec:PriceAccuracy}

The price shall reflect the true value of ETs. It shall be high enough to capture the highest possible share
of MEV while at the same time staying attractive for ticket holders to participate in the market.

As this is closely related to the objective of capturing MEV, we deem it feasible to measure it with the
same metrics proposed above for measuring the MEV capture.

\subsection{Ticket Parameters / Attributes}
\label{subsec:TicketParametersAttributes}

In order to structure the mechanism, it is essential to define certain parameters. In the following, we will
outline the most important design choices that need to be made and the possible configurations that can be
reasonably selected.

\subsubsection{Amount of Tickets}
\label{subsubsec:AmountOfTickets}

\paragraph*{Possible configuration: variable / fixed} 

The amount of tickets can either be set fixed at an arbitrary number or floating with a target amount. This will interplay with the pricing mechanism.

\subsubsection{Expiring Tickets}
\label{subsubsec:ExpiringTickets}

\paragraph*{Possible configuration: yes / no (if yes, expiry period)}

Tickets might be infinitely valid or for a fixed period of time. To ensure active participation in the lottery
and to avoid having “legacy” tickets around, it might be advantageous to have an expiration date.
However, this would complicate ticket pricing and their fungibility, since only tickets with the same
expiration date have the same value.

Another configuration could be to have batch expiring tickets (e.g. at the end of each quarter), however as
this only moves the pricing complications into the auction period (e.g. tickets sold on the 01/01 and 30/03
have different attributes), it is not further explored.

Further, if tickets are expiring a reasonable expiry period needs to be defined. Most likely, it makes sense
to choose this in relation to the total amount of outstanding tickets.

\subsubsection{Refundability}
\label{subsubsec:Refundability}

\paragraph*{Possible configuration: yes / no for allocated \& unallocated}

\citep{neuder_execution_2023} raised the question of whether or not ETs should be returnable to the protocol or not. The
advantage of refundability would be that this mechanism could reduce the number of outstanding tickets.
However, in addition to increased complexity, secondary questions arise, such as at what price can they be
returned. For example, shall they be repurchased at the current market price, the original purchase price or
any other price (e.g. a nominal small amount to allow to offload tickets)? If the price falls, will ticket
holders try to arbitrage by selling tickets to the protocol and buying them back directly?

\subsubsection{Resalability}
\label{subsubsec:Resalability}

\paragraph*{Possible configuration: yes / no for allocated \& unallocated}

Another question discussed was the topic whether tickets shall be resellable to third parties and a
secondary marketplace shall be allowed. This might be further differentiated between unallocated and
allocated tickets. resalability would increase the price of the tickets as it gives optionality and reduces risk
(e.g. what happens to ET holders that shut down their operations and cannot hold onto the promise of
validating a block?). \citep{neuder_execution_2023} proposed that an JIT-auction for allocated tickets with MEV-boost
may still exist. However, the resalability of allocated ETs could increase the risk of multi-round MEV and
financially potent actors extracting rent from owning several slots in a row if the lookahead period is
extended. If the lookahead period remains the same, the situation would remain at status quo where block
proposers can run MEV-Boost from the time they are scheduled to propose a block. As of today (October
2024) it has not been observed that block builders deliberately buy multi-slots for multi-round MEV,
however this might change in the future \citep{stichler_does_2024}. It generally opens the question if Execution
Tickets impose similar properties to slot auctions\footnote{The concept of slot auctions is introduced in more detail here:\newline \url{https://mirror.xyz/julianma.eth/CPYI91s98cp9zKFkanKs_qotYzw09kWvouaAa9GXBrQ}} only that the losses/profits are potentially captured by
Execution Ticket holders on the secondary market.

Another aspect to be considered in this, is how penalties can be introduced and enforced in case the (new)
ET holder misbehaves (e.g. misses a slot).

Next to protocol mechanism considerations the enforceability of the restriction however needs to be
further investigated. At our current knowledge, there are no credible mechanisms to prevent off-chain
reselling of slots. Hence, it might be more desirable to find an on-chain solution.

\subsubsection{Enhanced lookahead Period}
\label{subsubsec:EnhancedLookaheadPeriod}

\paragraph*{Possible configuration: no / yes ($x$ epochs)}

For the execution chain validators, the lookahead period may be increased to allow for better
predictability and better pre-confirmations. The current lookahead period for block proposers is set at one
epoch ahead (32 slots) and depends on the RANDAO\footnote{\url{https://github.com/ethereum/consensus-specs/blob/dev/specs/phase0/validator.md\#lookahead} \& \newline \url{https://docs.flashbots.net/flashbots-mev-boost/block-proposers}}. This means that practically as a maximum 64 slots
ahead are known (e.g. \citep{jensen_multi-block_2023}). A longer lookahead period on the execution chain might be
desirable, however might also increase the action space for multi-round MEV.

\subsection{Possible Pricing Mechanisms}
\label{subsec:PossiblePricingMechanisms}

Since Execution Ticket earnings are a proxy for block space similar properties to MEV-Boost emerge,
however with the significant difference that all rewards shall go to the protocol token holders by being
burned. There could be alternative ways to redistribute them, such as returning them to the users to reduce
transaction costs, however research shows that burning fees is often the mechanism that creates most
social welfare \citep{kiayias_would_2023}. Given the similar properties to block space, as \citep{buterin_blockchain_2019} has
shown, setting a fixed price or a fixed amount do not maximize social welfare. Hence, we will not
investigate pricing mechanisms that include a fixed price for assets, but will focus on dynamic pricing
mechanisms.

Generally, the pricing mechanisms can be divided into two categories of pricing tickets with either an
auction based format or an adaptive quoted price format. Furthermore, it can be structured with a variable
or fixed amount of tickets. As we observe that certain pricing mechanisms require specific configurations,
we will outline this for each pricing mechanism in the following.

\subsubsection{Pricing mechanisms with a quoted price}
\label{subsubsec:PricingMechanismsWithAQuotedPrice}

\paragraph{EIP-1559 style pricing}
\label{par:EIP1559StylePricing}

With the London hardfork on August 5, 2021 EIP-1559\footnote{\url{https://eips.ethereum.org/EIPS/eip-1559}} pricing for blockspace was introduced with a
base and priority fee. It was thoroughly investigated by e.g. \citep{roughgarden_transaction_2020} and found to be suitable
to decrease variance and increase user-experience. Further, \citep{roughgarden_transaction_2021} showed that from an
economic analysis in a restricted setting EIP-1559 is MMIC, usually DSIC and almost OCA-proof. For
Execution Tickets an adapted version of EIP-1559 pricing could be used as suggested by \citep{drake_session_2023}.
To maximize captured MEV the complete fee could be burned, which however might impair the
OCA-proofness. Process wise the protocol would quote a price for tickets and based on demand
(measured in the number of outstanding tickets compared to the desired amount of outstanding tickets) the
price would adapt similarly to EIP-1559. Hence, the number of outstanding tickets needs to be variable,
however with a specified target amount and a defined maximum adoption rate. for the beginning it might
make sense to use the 12.5\% from EIP-1559, however we test different adoption rates in the simulation.
Tickets could either be sold on a continuous basis or in a batch process where each slot between zero and
a specified maximum of tickets are sold.

In the last two years EIP-1559 has shown to be an efficient mechanism to keep the gas used close to the
target \citep{liu_empirical_2022}. Furthermore, the mechanism has been successfully implemented and is well
known making it an obvious choice. One of the main drawbacks is that the price adaption always only
happens retroactively and in times of MEV spikes might trail the actual market prices in the case of batch
processes. However, overall EIP-1559 seems like a promising pricing mechanism for a theoretical
perspective.

\paragraph{Bonding Curve / AMM style pricing}
\label{par:BondingCurveAMMStylePricing}

Automated market makers (AMMs) have been debated more intensely since a proposal by \citep{buterin_improving_2018}
to use them for DEXes. The basic idea is that two assets are deposited $x$ and $y$ with a defined function of $x
\ast y = k$, with $k$ being a constant. As demand changes the prices for the assets can move up and down the
bonding curve based on this. In \citep{buterin_make_2021-1} it is suggested making EIP-1559 more AMM-style to
have it more dynamically adapting by using an AMM-like price curve for blockspace. The proposal
includes a protocol parameter $excess\_gas\_issued$ and a pricing function:
\[ eth_{qty}(gas_{gty}) = exp\left( \dfrac{gas_{qty}}{gas_{Target}} \ast \dfrac{1}{\textit{adjustment Quotient}} \right) \]

It proposes that if a block proposer wants to make a block that contains $gas_in_block$ gas, they need to
pay 'burn fee' equal to $ eth_{qty} (\textit{excess gas issued} + \textit{gas in block}) - eth_{qty} (\textit{excess gas issued}) $.
After the block is processed the parameter is updated:
\[ \textit{excess gas issued} = \max(0, \textit{excess gas issued} + \textit{gas in block} - \textit{TARGET}) \]

While \citep{feist_exponential_2022} showed that the original version of EIP-1559 already has exponential properties, the
concept continues to be debated and e.g. for EIP-4844 a more exponential price adaptation mechanism
has been implemented\footnote{\url{https://github.com/ethereum/EIPs/blob/master/EIPS/eip-4844.md}}. For Execution Tickets, the mechanism can be adjusted accordingly with the
proposal for blockspace to allow for a passive price setting mechanism.

\begin{figure}[H]
    \centering
    \includegraphics[width=0.45\linewidth]{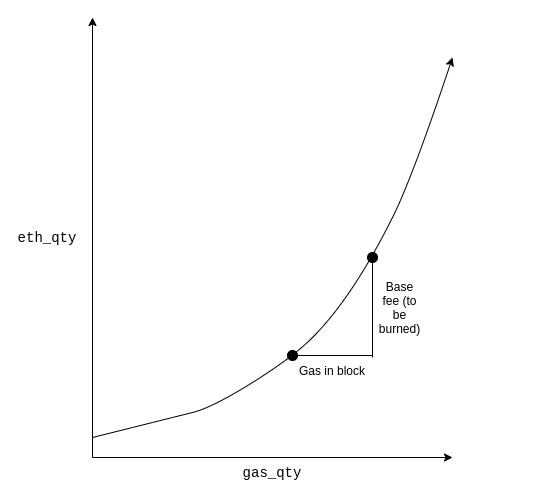}
    \caption{Exemplary pricing graph of bonding curve style pricing. Source: \citep{buterin_improving_2018}}
    \label{Fig_1}
\end{figure}

For Execution Tickets this concept can be adapted in several different manners. As AMMs are designed to
model the exchange rate between two assets it needs to be adapted for Execution Tickets, where one asset
can be bought at a certain price point and held in reference to a target amount of newly or currently issued tickets.

Three versions of how this can be adapted are outlined below.
\begin{enumerate}

    \item \textit{\textbf{EIP-1559 with AMM like curve:}} This approach most closely resembles the suggestion made in
    the Vitalik post mentioned above. In this case the price is updated after every slot. The parameter
    $excess\_gas\_issued$ is represented as $excess\_tickets\_issued = \max(0, excess\_tickets\_issued +
    tickets\_issued\_last\_slot - target\_tickets\_issued\_per\_slot)$. Target tickets issued per slot should be
    the same to the number of redeemed tickets per slot, usually being one. In summary it is an
    alternative price finding mechanism for EIP-1559 that shows more exponential behavior.
    
    \item \textit{\textbf{Continuous $\mathbf{e}$ delta update function:}} In this approach the ticket price is continuously updated
    after every ticket sale. However, to facilitate that for the regular EIP-1559 the \textit{gas target} is what
    should be consumed in total in the slot, for Execution Tickets the \textit{gas target} needs to be
    differentiated. In EIP-1559 no block proposer can 'hold' gas which concludes to different
    properties than holding tickets. For calculating the $excess\_tickets\_issued$ it should be the total of
    tickets that shall be held. For the quotient of the specific block it should be set to the number of
    tickets that shall be consumed in this slot (\textit{target tickets sold}), which in the normal setting is one
    to be equal to the number of redeemed tickets. The price of the ticket is calculated similar to the
    approach before as the delta between the two e functions as:
    $$\resizebox{0.93\linewidth}{!}{$
\textit{price(ticket)} = 
e^{\left( \dfrac{\textit{excess tickets held} + \textit{tickets sold}}{\textit{target tickets sold}} 
\ast \dfrac{1}{\textit{adjustment Quotient}} \right)} 
- 
e^{\left( \dfrac{\textit{excess tickets held}}{\textit{target tickets sold}} 
\ast \dfrac{1}{\textit{adjustment Quotient}} \right)}
$}$$
    
    With tickets sold being one as the price for the next ticket to be sold is calculated (for the next $n$
    tickets it would be $n$). This leads in a reduced function to:
    $$\resizebox{0.87\linewidth}{!}{$ price(ticket) = e^{\left((\textit{excess tickets held + 1}) \ast \dfrac{1}{\textit{adjustment Quotient}}\right)} - e^{\left((\textit{excess tickets held}) \ast \dfrac{1}{\textit{adjustment Quotient}}\right)} $}$$
    
    To denominate the expected amount of tickets in circulation it can be further adjusted by a
    constant $b$ to:
    \begin{align*}
    \textit{\small price(ticket)} &= \resizebox{0.78\linewidth}{!}{$e^{\left((\textit{excess tickets held + 1}) \ast \dfrac{1}{\textit{adjustment Quotient}} \right) + b } - e^{\left((\textit{excess tickets held}) \ast \dfrac{1}{\textit{adjustment Quotient}}\right) + b }$} \\
    &= \resizebox{0.78\linewidth}{!}{$e^b \ast \left( e^{(\textit{excess tickets held + 1}) \ast \dfrac{1}{\textit{adjustment Quotient}}} - e^{(\textit{excess tickets held}) \ast \dfrac{1}{\textit{adjustment Quotient}}} \right)$}
    \end{align*}
    
    This leads to the conclusion that the pricing function only depends on the \textit{excess tickets held} and
    the \textit{adjustment quotient}, being a constant. Hence, it is basically the delta between two e functions
    based on \textit{excess tickets held}.
    
    \item \textbf{\textit{Continuous $\bm{e}$ update function}}\footnote{If this concept is to be more widely debated, naming choices might need to be reconsidered.}\textbf{\textit{:}} Given the conclusion above a further approach might be to
    simplify the pricing function further to be simply dependent on the \textit{excess tickets held / tickets
    issued} without calculating the delta. This would result in the ticket price function:
    \[ price_{Ticket} (\textit{tickets issued}) = e^{b \ast \textit{tickets issued}} \] where $b$ is a constant defined by the protocol. 
    
    The constant $b$ could be calculated based on the desired \textit{TARGET AMOUNT} and \textit{TARGET PRICE} as:
    \[ b = \dfrac{\ln(\textit{TARGET PRICE})}{\textit{TARGET AMOUNT}} \]
    We will not investigate this approach further and leave it as an open area of research.

\end{enumerate}

\subsubsection{Auction-based Pricing mechanisms}
\label{subsubsec:AuctionBasedPricingMechanisms}

\paragraph{First price auction}
\label{par:FirstPriceAuction}

Sealed-bid first-price auctions (FPA) are auctions in which bidders submit their bids without knowing the
bids of others, and the highest bidder wins the item and pays the amount they bid \citep{nisan_best-response_2011}. A
major drawback of this auction format is the incentive for bid shading, where bidders may submit bids
lower than their true valuation to avoid paying too much, leading to inefficient outcomes where the item
does not necessarily go to the bidder who values it the most. This is formalized by \citep{roughgarden_transaction_2021}
for the fee market problem in the statement that FPAs are not DSIC.

This mechanism works best with a fixed amount of tickets, as the auction periods are ideally formalized.
Ideas about speeding up or slowing down the auction periods based on demand will not be further
investigated, as they add another axis of complexity while adding limited benefit. Demand regulation can
be done via price.

FPAs have the advantage of being a straightforward pricing mechanism and hence are used in the majority
of blockchain protocols currently \citep{ferreira_dynamic_2021}. On the downside they are as stated not DSIC, do
not reveal the true intrinsic willingness to pay of the market participants and might cause high volatility.

\paragraph{Second price auction}
\label{par:SecondPriceAuction}

Second-price auctions (SPA), also known as Vickrey auctions, are auctions where bidders submit their
bids without knowing the bids of others (sealed bids), and the highest bidder wins the item but pays the
second-highest bid. This auction format encourages bidders to disclose their true value as a bid, because
the winning bid only pays the amount of the second-highest bid \citep{nisan_best-response_2011}. However, a drawback
of second-price auctions is the potential for misbehavior of the auction holder to insert fake bids to drive
up the price. This is noted by \citep{roughgarden_transaction_2021} as not being MMIC. Given however, that the earnings
from Execution Tickets are not rewarded to the beacon chain validators this should pose less of a
problem. Further, \citep{roughgarden_transaction_2021} outlines that SPAs are almost OCA-proof. Overall, making SPAs
a promising pricing mechanism for Execution Tickets.

Operationally, several market participants have proposed how sealed bid auctions can be successfully
implemented in the Ethereum ecosystem (e.g. \citep{galal_verifiable_2018}).

\paragraph{Dutch auction}
\label{par:DutchAuction}

Dutch auctions start with a high starting price that gradually decreases until a bidder accepts the current
price and wins the item. This format is efficient for selling quickly and dynamically determining the
market price. The main drawback, however, is the potential for bidders to wait too long to bid, aiming to
get a lower price, which can lead to the item not being sold if the price decreases below the value bidders
are willing to pay. This results from the fact that they are not DSIC. This can be particularly problematic
in markets with fewer bidders or when the value of the item is not well-known, leading to uncertainty and
potentially lower revenues for the seller compared to other auction formats.

Further, in a protocol context it needs to be further investigated how latency might influence the bidding
dynamics and how the mechanism can be designed to avoid latency wars.

\paragraph{Other Auction Formats}
\label{par:OtherAuctionFormats}

There might be other more exotic pricing mechanisms such as the $\beta$-burn FPA proposed by
\citep{roughgarden_transaction_2021} or frequent batch auctions that we are currently not investigating further in the
scope of this work.

\subsubsection{Preliminary pricing mechanism evaluation}
\label{subsubsec:PreliminaryPricingMechanismEvaluation}

\citep{roughgarden_transaction_2021} analyzes six auction formats with regards to MMIC, DSCI and OCA-proofness in
the context of EIP-1559. It is shown that the EIP-1559 mechanism and the tipless mechanism (EIP-1559
style pricing without a tip) usually confine all three requirements. Only in a block with an excessively low
base fee they might fail to be DSIC / OCA-proof.
\begin{figure}[H]
    \centering
    \includegraphics[width=0.75\linewidth]{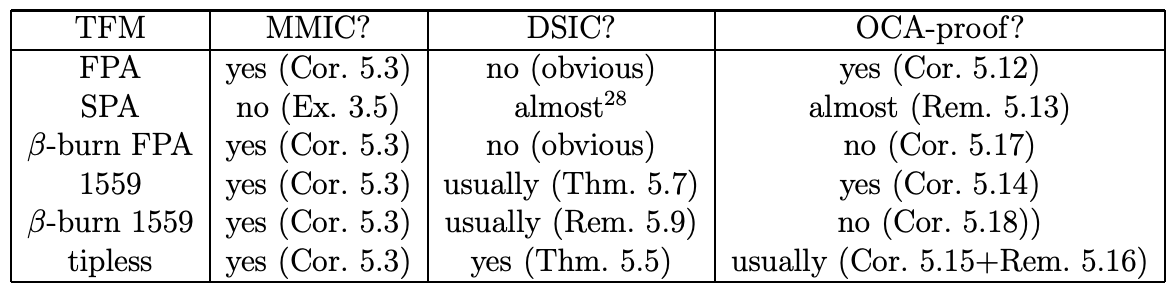}
    \caption{Overview of auction formats \citep{roughgarden_transaction_2021}}
    \label{Fig_2}
\end{figure}\vspace*{-8mm}

One limitation is that multi-round aspects and active block proposers are not considered. Additionally, if
rewards are burned, some of the mechanism dynamics change. Further, this analysis is limited to passive
block validators. \citep{bahrani_transaction_2023} show that it is fundamentally more difficult to achieve these
criteria with active block producers. In fact, they prove that no non-trivial or approximately
welfare-maximizing transaction fee mechanism can be incentive-compatible for both users and block
producers in the current setting.

Further, a winner's curse can occur in the auction formats, when the asset has a true value but different
actors have varying estimations of the true underlying value \citep{bazerman_i_1983}. By design
the most optimistic bidders will win the auction. Assuming a normal distribution of the estimations
around the true value, the highest bidders might be above the true value. This could be problematic in
cases when the variance among estimated values is high, as in turn it will either lead to optimistic market
participants leaving the market or generally adjusting their estimation by a variance factor.

Additionally as outlined in \citep{nisan_algorithmic_2007} under certain conditions all auction formats lead to the
same expected outcomes with the same revenues (Revenue Equivalence Principle). It is to note however,
that it assumes independent and identically distributed valuations for bidders. In the case of Execution
Ticket auctions, it can be assumed that private valuations of bidders are correlated as all bidders at least
partially base the valuation on the same available transactions visible in the public mempool.

\subsection{Target amount of tickets}
\label{subsec:TargetAmountOfTickets}

We deem it useful to define a target amount of circulating tickets, hence it needs to be controlled how
much tickets are sold in the primary market.

\citep{burian_execution_2024} has briefly touched on this. They show in their Theorem 4 that for a sufficiently large
number of tickets the current market cap of all tickets equals the present value of all future execution
layer rewards (MEV rewards plus fees).

Further, \citep{burian_execution_2024} outlined the trade-off between a smaller and larger amount of tickets also
summarized in \autoref{tab:SummaryOfAdvantagesAndDisadvantagesOfAHigherNumberOfTickets}. As postulated in Theorem 6 the expected value of a ticket decreases with the
amount of tickets. Further, as shown in Theorem 7 the value of commanding a share of the outstanding
tickets increases. Therefore, \citep{burian_execution_2024} argues that a large number of tickets protects against
monopolization as the cost of control increases. Further, a larger number decreases the cost per ticket and
thereby lowers barriers to entry. However, a larger number of tickets increases the complexity of ticket
pricing. In contrast, a smaller number of tickets allows for a shorter forecast horizon and hence potentially
less discounting and a higher capture of value. It however, results in a higher price per ticket. Further,
controlling a large percentage of the network would be less expensive as the total market capitalization of
issued tickets is lower.

\begin{table}[H]
\centering
\renewcommand{\arraystretch}{1.75} % Row height adjustment
\setlength{\arrayrulewidth}{0.15mm} % Thick lines
\setlength{\tabcolsep}{0.5em}

\adjustbox{max width = \linewidth}{
\begin{tabular}{|M{3.75cm}|Z{6cm}|Z{6cm}|}
    \hline
     & \textbf{Advantage} & \textbf{Disadvantage}
    \\ \hline

    \textbf{Larger \# of tickets}
    & 
    \begin{itemize}[left=0pt, topsep=5pt]
        \item Less centralization risk
        \item Lower cost of entry
    \end{itemize}
    &
    \begin{itemize}[left=0pt, topsep=5pt]
        \item Increased complexity of ticket valuation
        \item Potentially lower value capture
    \end{itemize}
    \\ \hline

    \textbf{Smaller \# of tickets}
    & 
    \begin{itemize}[left=0pt, topsep=5pt]
        \item Decreased complexity of ticket valuation
        \item Potentially higher value capture
    \end{itemize}
    &
    \begin{itemize}[left=0pt, topsep=5pt]
        \item Higher centralization risk
        \item Higher cost of entry
    \end{itemize}
    \\ \hline

\end{tabular}}

\vspace*{2mm}
\caption{Summary of advantages and disadvantages of a higher number of tickets}
\label{tab:SummaryOfAdvantagesAndDisadvantagesOfAHigherNumberOfTickets}

\end{table}

To ballpark the “sufficiently” large number of tickets defined in \citep{burian_execution_2024} we run a thought exercise
outlined in Appendix \hyperref[app:NumberOfExecutionTickets]{A}.

%% file: Sections/04-TheoreticalAnalysisOfSelectedMechanismDesigns.tex
\section{Theoretical Analysis of Selected Mechanism Designs}
\label{sec:TheoreticalAnalysisOfSelectedMechanismDesigns}

To substantiate the parameters in this section, several possible mechanism designs are discussed. Given
that based on the categorial parameters already 512 configurations are possible\footnote{$ 2 (Amount of Tickets) \ast 2 (Expiring Tickets) \ast 4 (Refundability) \ast 4 (Resalability)\ast 2 (Enhanced Lookahead) \ast 4 (Pricing Mechanisms)$}, only sample mechanism
designs are evaluated. Parameters are set differently and a theoretical evaluation of against the objective
parameters is given. In more detail, the following configurations are evaluated:

\begin{table}[H]
\centering
\renewcommand{\arraystretch}{1.25} % Row height adjustment
\setlength{\arrayrulewidth}{0.15mm} % Thick lines
\setlength{\tabcolsep}{0.5em}

\adjustbox{max width = \linewidth}{
\begin{tabular}{|M{2cm}|M{2.75cm}|M{2.75cm}|M{2.75cm}|M{2.75cm}|M{2.75cm}|M{2.75cm}|}

    \hline
    & \textbf{Simple FPA auction} & \textbf{JIT second price slot auction} & \textbf{Flexible 1559-style} & \textbf{Fixed SPA} & \textbf{Flexible, refundable AMM} & \textbf{Fixed, resellable FPA} \\ \hline
    Amount of tickets & Fixed & Fixed & Flexible & Fixed & Flexible & Fixed \\ \hline
    Expiring tickets & Yes & Yes & No & No & No & No \\ \hline
    Refundability & No (unallocated \& allocated) & No (unallocated \& allocated) & No (unallocated \& allocated) & No (unallocated \& allocated) & Yes (unallocated) & No (unallocated \& allocated) \\ \hline
    Resalability & No (unallocated \& allocated) & Yes (allocated) & Yes (unallocated \& allocated) & No (unallocated \& allocated) & No (unallocated \& allocated) & Yes (unallocated \& allocated) \\ \hline
    Enhanced Lookahead & No & Reduced & Yes for Execution Validators & Yes for Execution Validators & No & No \\ \hline
    \textbf{Pricing Mechanisms} & FPA & SPA & 1559-style & SPA & AMM & FPA \\ \hline
    \textbf{Target Amount} & 32 & 1 & undefined & 1024 & undefined & 1024 \\ \hline

\end{tabular}}

\vspace*{2mm}
\caption{Overview of possible mechanism design configurations}
\label{tab:OverviewOfPossibleMechanismDesignConfigurations}
\end{table}

\subsection{Simple FPA auction}
\label{subsec:SimpleFPAAuction}

To generate a baseline with regards to the evaluation of possible mechanism design configurations we will
first present a base strawman configuration against each other mechanism design can be benchmarked. It
will be designed as a straightforward auction of 32 execution tickets in the epoch $n - 1$ for the epoch $n$.

After the auction the tickets will be randomly assigned to slots. Note that this configuration has
similarities from the setup to the recently discussed mechanism Execution Auctions \citep{burian_execution_2024,elowsson_sealed_2024}. It will consist of the following mechanism design configuration:

\begin{table}[H]
\centering
\renewcommand{\arraystretch}{1.25} % Row height adjustment
\setlength{\arrayrulewidth}{0.15mm} % Thick lines
\setlength{\tabcolsep}{0.5em}

%\caption{}
%\label{}

\adjustbox{max width = \linewidth}{
\begin{tabular}{|Z{4cm}|Z{5cm}|}

\hline
\textbf{Ticket Attributes} & \textbf{Configurations} \\ \hline
Amount of tickets & Fixed \\ \hline
Expiring tickets & Yes (will be assigned in epoch $n$) \\ \hline
Refundability & No (unallocated \& allocated) \\ \hline
Resalability & No (unallocated \& allocated) \\ \hline
Enhanced Lookahead & No \\ \hline
\textbf{Pricing Mechanisms} & FPA \\ \hline
\textbf{Target Amount} & 32 \\ \hline

\end{tabular}}
\end{table}

\clearpage

With regards to the optimization objectives we can observe the following:

%\clearpage
\begin{table}[H]
\centering
\renewcommand{\arraystretch}{1.25} % Row height adjustment
\setlength{\arrayrulewidth}{0.15mm} % Thick lines
\setlength{\tabcolsep}{0.5em}

% \caption{}
% \label{}

\adjustbox{max width = \linewidth}{
\begin{tabular}{|Z{3cm}|Z{3cm}|Z{0.65\linewidth}|}

\hline
\multicolumn{3}{|l|}{\textbf{Simple FPA auction}} \\ \hline

\textbf{Objective} & \textbf{Est. Impact} & \textbf{Rationale} \\ \hline

Decentralization & \begin{enumerate}[label=\roman*)] \item High \item Low \end{enumerate} & \begin{enumerate}[label=\roman*)] \item Beacon chain validators: yes, in accordance with the ET rationale \item Execution chain validators: centralization forces due to specialization and ease of sophisticated actors to "game" FPAs. \end{enumerate} \\ \hline

MEV Capture & Medium & Given that FPAs are not DSIC and do not incentivize to bid the true value, the MEV capture will not be optimal\tablefootnote{Note that this assumes that the revenue equivalence principle does not apply as outlined in chapter \ref{subsubsec:PreliminaryPricingMechanismEvaluation}}. However, given the short lookahead period for the tickets, price predictability should be high and incentive to bid close to the expected value given enough competition. \\ \hline

Operational Process & Good & As FPAs are prevalent among protocols (former versions of Ethereum, Bitcoin etc.), operationalization should be good and feasible. \\ \hline

Price Predictability & Low & Given the short lookahead period and the auction mechanism, it is expected that prices have a high volatility and hence long-term predictability will be low. The high volatility is driven by the FPA, which as earlier versions of Ethereum pricing mechanisms have shown, allows for extreme volatility. \\ \hline

Price Smoothness & Low & See above. \\ \hline
Price Accuracy & Medium to High & The short lookahead period shall allow for a good price estimate of
validators, however the FPA does not incentive to reveal the
estimate and the FPA might cause high fluctuations \\ \hline

Other aspects & & n/a \\ \hline

\end{tabular}}
\end{table}

\subsection{JIT SPA slot auction}
\label{subsec:JIT_SPA_Slot_Auction}

\begin{table}[H]
\centering
\renewcommand{\arraystretch}{1.25} % Row height adjustment
\setlength{\arrayrulewidth}{0.15mm} % Thick lines
\setlength{\tabcolsep}{0.5em}

%\caption{}
%\label{}

\adjustbox{max width = \linewidth}{
\begin{tabular}{|Z{4cm}|Z{5cm}|}

\hline
\textbf{Ticket Attributes} & \textbf{Configurations} \\ \hline
Amount of tickets & Fixed at 1 \\ \hline
Expiring tickets & Yes  \\ \hline
Refundability & No (unallocated \& allocated) \\ \hline
Resalability & Yes (allocated) \\ \hline
Enhanced Lookahead & Reduced \\ \hline
\textbf{Pricing Mechanisms} & SPA \\ \hline
\textbf{Target Amount} & 1 \\ \hline

\end{tabular}}
\end{table}

As another strawman configuration we will investigate the maximal simple version of having a
just-in-time (JIT) second price slot auction. It can be seen as a special case of Execution Tickets, where
the number of tickets is restrained to one and this is sold just in time via a second price auction.
Alternatively also a first price auction could be considered, however as shown by e.g. \citep{varian_position_2007}
second price auctions are better in revealing the true willingness to pay. Further, the lookahead is set to
one, as only one ticket exists. The configuration has similarities to MEV-burn \citep{drake_mev_2023} however
might not require ePBS. Further, it has the properties of a slot auction not a block auction. It is similar
however to the design proposed \href{https://ethresear.ch/t/burning-mev-through-block-proposer-auctions/14029}{here} and \href{https://ethresear.ch/t/the-price-is-right-realigning-proposer-builder-incentives-with-predictive-mev-burn/18656}{here}.

This configuration can be seen as a strong simplification and extreme case which might reveal interesting
insights. What we can observe is that the market structure will be strongly simplified. Beacon chain
validators will propose the beacon block, the execution block will likely be proposed by the block
builders directly. For the block builders the market dynamics will look similar, they create profitable
blocks and then instead of bidding in MEV-Boost they will bid in the new JIT auction. Timing and setting
the right timing windows will be important in this structure. Given a long enough time in between, it
could be that a separate entity is buying the execution ticket and then auctioning the block building off to
block builders. If the timing is set short enough similar to the current MEV-Boost we assume that MEV
capture will be very high as block builders have the maximum amount of information available to make a
value capturing bid. From an operational process the setup might be challenging given the short time
horizons. Especially, reliably defining the winning bid and communicating it might be a challenge. Price
predictability and smoothness is estimated to be low as they will move in accordance with the current
MEV reward curve. Price accuracy shall be very high, as the bidders have the maximum amount of
information possible and only carry a low risk of a ticket holding period, given the short holding period.

In conclusion, we observe that MEV capture and decentralization are sufficiently high in this scenario,
however from an operational perspective it might not be feasible as defining the winning bid could pose a
challenge and a low price predictability and smoothness might occur.

\begin{table}[H]
\centering
\renewcommand{\arraystretch}{1.25} % Row height adjustment
\setlength{\arrayrulewidth}{0.15mm} % Thick lines
\setlength{\tabcolsep}{0.5em}

% \caption{}
% \label{}

\adjustbox{max width = \linewidth}{
\begin{tabular}{|Z{3cm}|Z{3cm}|Z{0.65\linewidth}|}

\hline
\multicolumn{3}{|l|}{\textbf{JIT SPA slot auction}} \\ \hline

\textbf{Objective} & \textbf{Est. Impact} & \textbf{Rationale} \\ \hline

Decentralization & \begin{enumerate}[label=\roman*)] \item High \item Medium \end{enumerate} & \begin{enumerate}[label=\roman*)] \item Beacon chain validators: yes, in accordance with the ET rationale \item Execution chain proposers: expected to be similar to the current builder market, as for block builders the market will look the same. \end{enumerate} \\ \hline

MEV Capture & High & Maximum information available to short-term predict block value \\ \hline

Operational Process & Difficult & Short time window increases complexity. \\ \hline

Price Predictability & Low & Long term price predictability will be low, given that the price will closely follow the real amount of MEV and fees generated \\ \hline

Price Smoothness & Low & As MEV is spikey, so will be the ticket prices. \\ \hline

Price Accuracy & High & The short time horizon and pricing mechanism will allow for a good price accuracy. \\ \hline

Other aspects & &  \\ \hline

\end{tabular}}
\end{table}

\subsection{Flexible 1559-style Configuration}
\label{subsec:Flexible1559StyleConfiguration}

As another configuration we will evaluate a configuration inspired by the proposal from \citep{neuder_execution_2023}.
Based on the proposal we construct a configuration drawing on several assumptions. ETs are sold using
1559-style pricing where the price is adjusted based on a targeted total supply of tickets in circulation. At
what frequency the tickets are being sold is flexible. Hence the amount of outstanding tickets is flexible
with a predefined target supply. Tickets are set to be non-expiring. The same goes for refundability.
Tickets are designed to be resellable. The lookahead period for beacon chain validators is short to reduce
timing games, however longer for execution chain validators to allow for pre-confirmations.

\begin{table}[H]
\centering
\renewcommand{\arraystretch}{1.25} % Row height adjustment
\setlength{\arrayrulewidth}{0.15mm} % Thick lines
\setlength{\tabcolsep}{0.5em}

%\caption{}
%\label{}

\adjustbox{max width = \linewidth}{
\begin{tabular}{|Z{4cm}|Z{5cm}|}

\hline
\textbf{Ticket Attributes} & \textbf{Configurations} \\ \hline
Amount of tickets & Fixed \\ \hline
Expiring tickets & No  \\ \hline
Refundability & No (unallocated \& allocated) \\ \hline
Resalability & Yes (unallocated \& allocated) \\ \hline
Enhanced Lookahead & Yes for Execution Validators \\ \hline
\textbf{Pricing Mechanisms} & 1559-style \\ \hline
\textbf{Target Amount} & undefined \\ \hline

\end{tabular}}
\end{table}

\clearpage

Preliminarily we can see that this results in a high decentralization among beacon chain validators and a
medium decentralization amongst execution ticket holders. The general centralization forces on the
execution chain validators persist. However, new market entrants can always join the market and if they
are not technically sophisticated enough for block building can auction off their ticket / slot. MEV capture
is assumed to be good, given that 1559-style pricing adjusts dynamically to the demand. It is crucial
therefore however that the target amount of tickets, lower and upper bound as well as adjustment
mechanism is specified well.

\begin{table}[H]
\centering
\renewcommand{\arraystretch}{1.25} % Row height adjustment
\setlength{\arrayrulewidth}{0.15mm} % Thick lines
\setlength{\tabcolsep}{0.5em}

% \caption{}
% \label{}

\adjustbox{max width = \linewidth}{
\begin{tabular}{|Z{3cm}|Z{3cm}|Z{0.65\linewidth}|}

\hline
\multicolumn{3}{|l|}{\textbf{Flexible 1559-style Configuration}} \\ \hline

\textbf{Objective} & \textbf{Est. Impact} & \textbf{Rationale} \\ \hline

Decentralization & \begin{enumerate}[label=\roman*)] \item High \item Medium \end{enumerate} & \begin{enumerate}[label=\roman*)] \item Beacon chain validators: yes, in accordance with the ET rationale \item Execution chain validators: general centralization forces due to specialization, however 1559-style pricing always allows for new market entrants. Given the resalability less sophisticated actors can always auction off their assigned slot. \end{enumerate} \\ \hline

MEV Capture & Medium to Good & Assuming a suitable target amount 1559-style pricing should allow for an efficient price discovery to adjust the price to the demand, however with a delay of 1-slot. \\ \hline

Operational Process & Good & As 1559-style pricing is currently implemented for block space the concept is known to actors. \\ \hline

Price Predictability & Very low & Short term predictability is good, given the incremental price adjustment mechanism. However, long term is harder to forecast as the number of tickets varies. \\ \hline

Price Smoothness & Medium to High & Given the price adjustment mechanism lower variance is expected. \\ \hline

Price Accuracy & Medium to High & Accuracy depends on the number of circulating tickets and the expected internal pricing mechanism. \\ \hline

Other aspects & & Multi-MEV is possible to the same extent it is currently, as block builders could buy multi-slots on the secondary market via the JIT-auctions of block space of execution chain validators. \\ \hline

\end{tabular}}
\end{table}

\subsection{Fixed SPA}
\label{subsec:FixedSPA}

Another configuration we will investigate is a sealed-bid fixed second price auction (SPA), with a fixed
amount of tickets of an arbitrary number of 1024\footnote{Note that for performance limitations, less tickets are chosen in the simulation} and restrained resalability. In this setting, there needs
to be an allocation mechanism for the initial 1024 tickets and then each slot, one ticket is sold and
redeemed, keeping the number of outstanding tickets fixed. It is to note, that it has yet to be investigated
how to operationally restrict the secondary market.

\begin{table}[H]
\centering
\renewcommand{\arraystretch}{1.25} % Row height adjustment
\setlength{\arrayrulewidth}{0.15mm} % Thick lines
\setlength{\tabcolsep}{0.5em}

%\caption{}
%\label{}

\adjustbox{max width = \linewidth}{
\begin{tabular}{|Z{4cm}|Z{5cm}|}

\hline
\textbf{Ticket Attributes} & \textbf{Configurations} \\ \hline
Amount of tickets & Fixed \\ \hline
Expiring tickets & No  \\ \hline
Refundability & No (unallocated \& allocated) \\ \hline
Resalability & No (unallocated \& allocated) \\ \hline
Enhanced Lookahead & Yes for Execution Validators \\ \hline
\textbf{Pricing Mechanisms} & Second price auction \\ \hline
\textbf{Target Amount} & 1024 \\ \hline

\end{tabular}}
\end{table}

\clearpage
In this setting, tickets for the next 32 epochs are outstanding spanning roughly the next 3.5 hours. In
accordance with \citep{burian_execution_2024} this means the expected time until a ticket is selected is as well 1024
slots. This gives a short forecasting range and allows for accurate ticket pricing. The beacon chain
validators are expected to be highly decentralized in accordance with basic ET assumption. Execution
chain proposers are expected to be less decentralized as higher ticket prices and a limited number of
tickets will restrict the supply. In the current market as outlined in Chapter \ref{subsec:TicketParametersAttributes}, ticket prices can be
estimated to be close to the expected MEV of 0.05 ETH (\$175) based on theorem 2 by \citep{burian_mev_2024}
where $d$ can be neglected. This would result in a low total market capitalization of outstanding tickets of
51.2 ETH (\$179.200). However, it will be easy for new market entrants to purchase a meaningful amount
of outstanding tickets, given the limited supply.

\begin{table}[H]
\centering
\renewcommand{\arraystretch}{1.25} % Row height adjustment
\setlength{\arrayrulewidth}{0.15mm} % Thick lines
\setlength{\tabcolsep}{0.5em}

% \caption{}
% \label{}

\adjustbox{max width = \linewidth}{
\begin{tabular}{|Z{3cm}|Z{3cm}|Z{0.65\linewidth}|}

\hline
\multicolumn{3}{|l|}{\textbf{Fixed SPA}} \\ \hline

\textbf{Objective} & \textbf{Est. Impact} & \textbf{Rationale} \\ \hline

Decentralization & \begin{enumerate}[label=\roman*)] \item High \item Low to Medium \end{enumerate} & \begin{enumerate}[label=\roman*)] \item Beacon chain validators: yes, in accordance with the ET rationale \item Execution chain proposer: general centralization forces due to specialization and higher prices per ticket. However, easy for new market entrants to start participating and purchase tickets, however they need the infrastructure to propose blocks as no resalability possible. \end{enumerate} \\ \hline

MEV Capture & Good & Given that SPAs are a very efficient pricing mechanism and the uncertainty with a short redemption period is low, MEV capture is expected to be very high. Only the missing resalability might limit the MEV capture. \\ \hline

Operational Process & Good & Assuming that leaderless sealed-bid auctions can be implemented, operational overhead should be limited. \\ \hline

Price Predictability & Low & Hard to forecast, as spikes in MEV might directly influence the pricing of tickets. \\ \hline

Price Smoothness & Low & Given the auction format, price spikes can occur. \\ \hline

Price Accuracy & High & Short uncertainty period and accurate pricing mechanism. \\ \hline

Other aspects & & Limited number of tickets makes it easy for one actor to purchase a majority of the tickets (e.g. owning >51\% can be estimated at an investment of $\sim\$90.000$). However, randomization might limit the extent multi-block MEV is possible. \\ \hline

\end{tabular}}
\end{table}

\subsection{Flexible, refundable AMM}
\label{subsec:FlexibleRefundableAMM}

Another configuration to be investigated is an AMM-like mechanism with a flexible amount of
refundable tickets. A bonding curve is defined which indicates the price and allows market participants to
buy or sell tickets to at any time. The refundability is restrained to unallocated tickets. Resalability is
constrained in this scenario and the target amount is undefined, as the amount of outstanding tickets
interplays with the price.

\begin{table}[H]
\centering
\renewcommand{\arraystretch}{1.25} % Row height adjustment
\setlength{\arrayrulewidth}{0.15mm} % Thick lines
\setlength{\tabcolsep}{0.5em}

%\caption{}
%\label{}

\adjustbox{max width = \linewidth}{
\begin{tabular}{|Z{4cm}|Z{5cm}|}

\hline
\textbf{Ticket Attributes} & \textbf{Configurations} \\ \hline
Amount of tickets & Flexible \\ \hline
Expiring tickets & No  \\ \hline
Refundability & Yes (unallocated) \\ \hline
Resalability & No (unallocated \& allocated) \\ \hline
Enhanced Lookahead & No \\ \hline
\textbf{Pricing Mechanisms} & AMM \\ \hline
\textbf{Target Amount} & undefined \\ \hline

\end{tabular}}
\end{table}

\clearpage
Decentralization effects are expected to be similar to other configurations. With regard to MEV capture an
AMM-style pricing might be good at gauging the true intrinsic ticket valuation.

\begin{table}[H]
\centering
\renewcommand{\arraystretch}{1.25} % Row height adjustment
\setlength{\arrayrulewidth}{0.15mm} % Thick lines
\setlength{\tabcolsep}{0.5em}

% \caption{}
% \label{}

\adjustbox{max width = \linewidth}{
\begin{tabular}{|Z{3cm}|Z{3cm}|Z{0.65\linewidth}|}

\hline
\multicolumn{3}{|l|}{\textbf{Flexible, refundable AMM}} \\ \hline

\textbf{Objective} & \textbf{Est. Impact} & \textbf{Rationale} \\ \hline

Decentralization & \begin{enumerate}[label=\roman*)] \item High \item Low to Medium \end{enumerate} & \begin{enumerate}[label=\roman*)] \item Beacon chain validators: yes, in accordance with the ET rationale \item Execution chain proposer: general centralization forces due to specialization and potentially high ticket prices. Other effects are hard to predict. \end{enumerate} \\ \hline

MEV Capture & High & Given that ET buyers are going low risk with the opportunity to return tickets to the protocol they can buy very close to the true value and the price will dynamically adjust. \\ \hline

Operational Process & Good & Given that trustless implementations for AMMs exist it should be feasible to adapt. \\ \hline

Price Predictability & Low & Hard to forecast, as spikes in MEV might directly influence the pricing of tickets. \\ \hline

Price Smoothness & Low to Medium & Prices can adapt quickly, however in a continuous price movement. \\ \hline

Price Accuracy & High & Adaptive pricing mechanism that also takes into account
sell-pressure in the market. \\ \hline

Other aspects & & At least a few market participants are needed for effective price setting, as in all other configurations. \\ \hline

\end{tabular}}
\end{table}

\subsection{Fixed, resellable FPA}
\label{subsec:FixedResellableFPA}

The sixth tested configuration is a first-price auction with a fixed amount of tickets and a secondary
market enabled.

\begin{table}[H]
\centering
\renewcommand{\arraystretch}{1.25} % Row height adjustment
\setlength{\arrayrulewidth}{0.15mm} % Thick lines
\setlength{\tabcolsep}{0.5em}

%\caption{}
%\label{}

\adjustbox{max width = \linewidth}{
\begin{tabular}{|Z{4cm}|Z{5cm}|}

\hline
\textbf{Ticket Attributes} & \textbf{Configurations} \\ \hline
Amount of tickets & Fixed \\ \hline
Expiring tickets & No  \\ \hline
Refundability & No (unallocated \& allocated) \\ \hline
Resalability & Yes (unallocated \& allocated) \\ \hline
Enhanced Lookahead & No \\ \hline
\textbf{Pricing Mechanisms} & FPA \\ \hline
\textbf{Target Amount} & 1024 \\ \hline

\end{tabular}}
\end{table}

Decentralization effects are expected to be similarly low as other configurations with a longer lookahead
on the primary market however might increase on the JIT secondary market. With regard to MEV capture
it highly depends on the bidding strategies and competitiveness of the market.

\clearpage

\begin{table}[H]
\centering
\renewcommand{\arraystretch}{1.25} % Row height adjustment
\setlength{\arrayrulewidth}{0.15mm} % Thick lines
\setlength{\tabcolsep}{0.5em}

% \caption{}
% \label{}

\adjustbox{max width = \linewidth}{
\begin{tabular}{|Z{3cm}|Z{3cm}|Z{0.65\linewidth}|}

\hline
\multicolumn{3}{|l|}{\textbf{Fixed, resellable FPA}} \\ \hline

\textbf{Objective} & \textbf{Est. Impact} & \textbf{Rationale} \\ \hline

Decentralization & \begin{enumerate}[label=\roman*)] \item High \item Low to Medium \end{enumerate} & \begin{enumerate}[label=\roman*)] \item Beacon chain validators: yes, in accordance with the ET rationale \item Execution chain proposer: Low decentralization on the primary market, however specialization might drive more decentralization on the secondary market \end{enumerate} \\ \hline

MEV Capture & Medium to High & Given that FPAs are not DSIC and do not incentivize to bid the true value, the MEV capture will not be optimal. However, depending on the competitiveness of the bidding might still be high. \\ \hline

Operational Process & Good & As FPAs are prevalent among protocols (former versions of Ethereum, Bitcoin etc.), operationalization should be good and feasible. \\ \hline

Price Predictability & High & Given that bidding will be done on expected values, bids should remain stable and predictable. \\ \hline

Price Smoothness & High & Given that bidding will be done on expected values, bids should remain stable. \\ \hline

Price Accuracy & Medium to High & Accuracy depends on competitiveness of FPA and bidding strategies. \\ \hline

Other aspects & & Remains open if sealed-bid first price auctions could be pragmatically implemented in this setting. \\ \hline

\end{tabular}}
\end{table}

%% file: Sections/05-Simulation.tex
\section{Simulation}
\label{sec:Simulation}

\subsection{Simulation Setup}
\label{subsec:Simulation Setup}

For the simulation, we have set up a framework to test how the process around execution tickets looks
like with several variables that can be adjusted to test holistically different configurations. It is set up in a
way that other researchers can dynamically test other configurations. For the purpose of this study, we
selected several mechanism design configurations and stress tested them with agent based-modeling. This
provided more in-depth insights regarding the ideal parameter configurations and what are the effects of
each individual parameter. Furthermore, cross effects became visible and can be better investigated. For
each configuration it is analyzed how well it performs with regards to the defined objectives and
respective metrics.

To do this, we define a simulation of the Execution Ticket issuance, trading and redemption process. This
is done using \href{https://github.com/CADLabs/radCAD/tree/master?tab=readme-ov-file}{radCAD}. The framework is designed to model different settings for the selling process of
tickets as well as adjusting the ticket holders. The source code for the simulation can be found under \href{https://github.com/ephema/ET-Simulation}{this
link}.

As common for radCAD/cadCAD simulations, it is divided into policy functions and variable update
functions.

The simulation is defined as:
\begin{lstlisting}[language=Python]
simulation = Simulation(model=model, timesteps=TIMESTEPS, runs=RUNS)
\end{lstlisting}

Per default the \textit{TIMESTEPS} and \textit{RUNS} are set to 1000 steps (slots) and 10 runs, this can be adjusted. One
timestep represents one slot on the blockchain.

The simulation model is defined as:
\begin{lstlisting}[language=Python]
model = Model(
    params=sys_params,
    initial_state=initial_state,
    state_update_blocks=state_update_blocks)
\end{lstlisting}

The params for the simulation hold the parameters to be tested. It includes: '\textit{selling\_mechanism}',
'\textit{max\_tickets}', '\textit{initial\_ticket\_price}', '\textit{MEV\_scale}', '\textit{slots\_per\_epoch}',  '\textit{number\_of\_ticket\_holders}', '\textit{secondary\_market}', \newline'\textit{price\_vola}', '\textit{agent\_bidding\_strategy}', '\textit{EIP-1559\_max\_tickets}', '\textit{EIP-1559\_adjust\_factor}',  \newline'\textit{AMM\_adjust\_factor}', '\textit{expiry\_period}' 
and '\textit{reimbursement\_factor}'.

The \textit{initial\_state} is calculated in the beginning and involves generating the different entities (agents,
tickets etc.) and generating and allocating a first batch of tickets.

The \textit{state\_update\_blocks} are separated into four sub-steps defined as:
\begin{enumerate}
    \item \textbf{\textit{Update Market Meta Data:}} New tickets are issued, slot count is increased, MEV for this slot is defined etc.
    \item \textbf{\textit{Purchase Tickets:}} Tickets can be purchased by ticket holders according to the defined mechanism.
    \item \textbf{\textit{Secondary Market:}} If enabled, the secondary market is run.
    \item \textbf{\textit{Redeem Tickets:}} Tickets are redeemed by ticket holders, MEV captured is updated etc.
\end{enumerate}

\subsubsection{Specification}
\label{subsubsec:Specification}

For the simulation we need to define some variables and entities.

\paragraph{Simulation Environment}
\label{par:SimulationEnvironment}

To provide the necessary metadata and monitor the following environment variables are set up: \textit{slot,
epoch}, \textit{current\_ticket\_id}, \textit{MEV\_per\_slot}, \textit{Volatility\_per\_slot} and \textit{total\_MEV\_captured}.

\textbf{\textit{Available MEV per slot:}} It is assumed that per slot there is a theoretical amount of MEV available that
can be captured. However, the true value is not known to market participants. The MEV is generated in
each slot and follows an exponential distribution (as observed in historical data, see e.g. \citep{stichler_does_2024}).
It can be seeded with the constant \textit{MEV\_scale}. The historical MEV values are known to market
participants including the average and distribution. So this can be used by ticket holders for bidding
mechanisms.

\textbf{\textit{Volatility per slot:}} for each slot a randomized token price volatility is generated and it can be enabled that
volatility interplays with the specialization of builders. See below for details.

\paragraph{Execution Ticket}
\label{par:ExecutionTicket}

Execution tickets are batch-allocated in step 0 of the simulation and then behave according to multiple
adjustable parameters (described in Chapter \ref{sec:MechanismObjectivesAndDesignSpace}):
\begin{itemize}[itemsep = -2mm]
    \item Amount of tickets (variable vs. fixed)
    \item expiring ticket (yes/no)
    \item refundable
    \item resellable
    \item target amount of tickets \& pricing mechanism
\end{itemize}

Further, tickets need to have the following attributes:
\begin{itemize}[itemsep = -2mm]
    \item assigned vs. unassigned
    \item assigned slot
    \item assigned epoch
    \item redeemed slot
    \item redeemed epoch
    \item Expiry slot (if enabled)
\end{itemize}

\paragraph{Execution Ticket Holder}
\label{par:ExecutionTicketHolder}

Execution ticket holders have several attributes:
\begin{itemize}[itemsep = -2mm]
    \item Type (see below)
    \item Number of tickets in possession
    \item ability to capture MEV (in percentage)
    \item intrinsic ticket valuation (outlining the expectation of the ticket holder on the ticket's worth at
    \item redemption)
    \item discount factor
    \item list of tickets in possession
    \item accumulated earnings
    \item accumulated costs
    \item available funds (tbd)
    \item aggressiveness in bidding (required margin on tickets)
    \item Volatility specialization factor
\end{itemize}

Execution ticket holders are buying tickets on the primary market and are able to buy and sell tickets on
the secondary markets in applicable cases.

In accordance with the findings of \citep{yang_decentralization_2024} the builder market is expected to be split into the
three groups: top (20\%), middle (40\%) and tail (40\%) builders. As the number of builders can be
manually adjusted in the experiment parameters, the split is implemented as a relative split. It is assumed
that the MEV extraction capabilities as well as available funds vary between the groups but intra-groups
are similar. Hence, in the implementation the \textit{MEV\_capture\_rate} as well as \textit{available\_funds} is initialized
as a random uniform distribution for each group separately with a small interval window to reflect that the
abilities are similar.

The ability to capture MEV in percentage shows the proficiency of execution ticket holders to extract a
percentage of the “available MEV per slot”. Sophisticated actors are able to extract higher percentages of
the available MEV.

The intrinsic ticket valuation guides the demand function of ticket holders and is based on the historic
experience of capturable MEV (available MEV times ability to capture MEV). Rational ticket holders are
expected to use the pricing function outlined in Chapter \ref{subsubsec:BiddingBehavior2}. to purchase tickets. In auctions they will
bid value maximizing (delta between price and discounted intrinsic ticket valuation times winning
probability). For quoted price mechanisms (EIP-1559 style and AMM-style) they will buy a ticket if the
price is below their intrinsic valuation. For the simulation several different configurations will be tested of
execution ticket holders with similar abilities to capture MEV as well as more diverse abilities.

The default Execution Ticket holder seeding is generated as:
\begin{lstlisting}[language=Python]
if self.id <= (num_holders * 0.2):
    self.type = 'top'
    self.available_funds = random.uniform(400, 1000)
    self.MEV_capture_rate = random.uniform(0.85, 0.95)
elif self.id > (num_holders * 0.2) and self.id <= (num_holders * 0.6):
    self.type = 'middle'
    self.available_funds = random.uniform(300, 700)
    self.MEV_capture_rate = random.uniform(0.75, 0.85)
elif self.id > (num_holders * 0.6):
    self.type = 'tail'
    self.available_funds = random.uniform(200, 500)
    self.MEV_capture_rate = random.uniform(0.6, 0.75)
self.aggressiveness = np.random.normal(0.15, 0.02)
self.vola_spec_factor = np.random.normal(1, 0.5)
if self.vola_spec_factor <= 0: self.vola_spec_factor = 0.1
\end{lstlisting}

\paragraph{Secondary market}
\label{par:SecondaryMarket}

To realistically model the implementation of Execution Tickets a secondary market place is implemented
and can be toggled on. As outlined earlier it is unclear if a secondary market could even be technically
prevented. As \href{https://boost.flashbots.net/}{MEV-Boost} is currently the dominant implementation for a secondary block market, we
orient the simulation on it. MEV-Boost is implemented as a public first price auction, which essentially
results in a second price auction. Hence, we implement the secondary market in the implementation as a
second price auction.

In each round, each ticket holder has the opportunity to offer their tickets for sale to all other ticket
holders. For brevity we limit the amount of tickets per holder to be sold per round to one. In this context it
is important to note that tickets might be non-fungible assets. This derives from the specific attributes of
the ticket: \textit{expiry\_slot}, \textit{assigned\_slot} and \textit{volatility\_per\_slot}. In case the configuration is set up with
non-expiring tickets, no discount factor for the assigned slot and the volatility for the slot is unknown the
ticket is fungible. In all other cases tickets are non-fungible. Hence, these factors need to be included in
the pricing function, when ticket holders decide on their bids for the offered tickets. The pricing for the
secondary market can be derived from the pricing for the second price auction described below.

Additionally, the pricing for expiring tickets and assigned slots (discount factor and volatility) needs to be
included in the pricing functions.

The basic pricing for expiring tickets is described below. In the case of secondary market tickets, the
variable S noting the total slots a ticket is valid needs to be adapted to S' denoting the remaining period a
ticket is valid (e.g. if a ticket was issued in slot 100 and is valid for 50 slots and is sold in slot 110, the
remaining period is 40). Furthermore, it needs to be checked if the ticket is allocated to a slot, in this case
the expiry-risk is zero. Additionally, it could be taken into consideration, how many of the future slots are
already allocated to other tickets (e.g. how long the lookahead is), for runtime efficiency reasons we
currently do not include this in the calculation, however could be simply done by iterating over the tickets
and collecting the assigned slots and dividing these by the outstanding tickets.

The pricing for volatility is a simple adjustment for the \textit{vola\_spec\_factor} of each ticket holder, if the
volatility is known. In the current implementation this is only relevant for same slot ticket sales.

\paragraph{Observation Parameters}
\label{par:ObservationParameters}

To monitor the simulation outcomes we define a few observation parameters that we use to evaluate each
configuration:
\begin{itemize}[itemsep = -2mm]
    \item Market Share of all Execution Ticket holders
    \item Nakamoto-coefficient
    \item Herfindahl-Hirschman Index
    \item MEV-Share Protocol
    \item Garman-Klass (GK) Measure
    \item Variance of deltas ($V(\Delta p)$)
\end{itemize}

\subsubsection{Pricing Mechanism}
\label{subsubsec:PricingMechanism}

For the simulation four pricing mechanisms are implemented: first price auction (FPA), second price
auction (SPA), EIP-1559 style pricing and AMM-style pricing.

As presented above for EIP-1559 it could either be defined as a continuous process or as a batch process.
We chose to implement it as a batch process to make it more aligned with the original EIP-1559 pricing
that updates after every slot. Further for the EIP-1559 style pricing two parameters are defined that can be
adjusted in the \textit{sys\_params}: \textit{EIP-1559\_max\_tickets} and \textit{EIP-1559\_adjust\_factor}. The first one defines
how many people per round as a maximum can be purchased by ticket holders. The second defines the
price adjustment factor. This is multiplied by the relative delta between outstanding tickets and target
amount of tickets (e.g. if 10\% more tickets are outstanding, the price is adjusted by $ 0.1 \ast (1/EIP - 1559\_adjust\_factor) $. 
We start with a factor of 8, to represent the 12.5\% of the 
EIP-1559 pricing and adjust this in subsequent simulations.

The AMM-style pricing is implemented as an alternative price finding mechanism for the EIP-1559
mechanism, where the pricing is implemented in accordance with the second suggestion labeled
“Continuous e delta update function”. For the \textit{adjustment quotient} the constant \textit{AMM\_adjust\_factor} is
defined and set to 6 in the base case. A more formal way of defining the right adjustment factor here
remains the scope of future research. Also for further research we believe the AMM-style pricing shall be
further refined before it is in a finalized version.

\subsubsection{Bidding Behavior}
\label{subsubsec:BiddingBehavior2}

For the ticket holder agents different bidding strategies are implemented and compared starting from very
simple bidding strategies to increasingly sophisticated bidding. Obviously the bidding strategies heavily
depend on the information the ticket holders have at hand. We generally assume that the ticket holders
have access to historical MEV values and can use them as a baseline for predicting future values. Further,
we assume there is no median drift and the historic data is indeed a good baseline for prediction. Further,
we assume that bids are submitted as sealed-bids and not visible to other market participants. The bidding
strategies depend on the selling mechanism and are in more detail:

\paragraph{First price Auction}
\label{par:FirstPriceAuction2}

For first price auctions we implemented several different strategies starting with a random even
distribution around the historical median value. Further, we implemented a naive observation of historical
true MEV values, which are then multiplied by the aggressiveness (margin requirement) of the agent. In
the next strategy the ticket holder is aware of their ability to capture MEV, hence this is factored into the
bidding.

Further, as noted in literature an approximal optimal bidding strategy for sealed-bid first price auction is
to bid \( \frac{n - 1}{n} \ast \textit{intrinsic valuation} \) with $n$ denoting the number of other active bidders \citep{greenwald_first-price_2017}. This is implemented as well with builders being aware of the number of market
participants and adjusting their bid accordingly. A next step of adaptive bidding would be to observe how
many other bidders have submitted competitive bids in the last $n$ slots and use this as an approximation
for $n$. We have not implemented this, as we only expected incremental results but could be easily adapted.

\paragraph{Second price Auction}
\label{par:SecondPriceAuction2}

For second price auctions we implemented the same strategies following the logic for the intrinsic value
from the first price auction, only with the adjustment that the true intrinsic value is submitted as bids as
this is the dominant strategy for second price auctions.

In case the auction is at the same slot as the ticket (for first or second price auction) the ticket holders are
aware of the volatility in this slot and this is included in the bidding strategy.

\paragraph{AMM-Style pricing}
\label{par:AMM_StylePricing}

For the AMM-style mechanism an intrinsic valuation is calculated based on the historical MEV scale, the
MEV capture rate, the aggressiveness and if applicable the discount factor for expiring tickets.

\paragraph{EIP-1559-style Pricing}
\label{par:EIP1559StylePricing2}

The bidding strategy for the EIP-1559-style mechanism is analogous to the AMM-style pricing.

\subsubsection{Expiring Tickets}
\label{subsubsec:ExpiringTickets2}

Expiring tickets are implemented as a property. One decision to make is how to deal with the initial
allocation of tickets. In the implementation they are set to be expiring linearly from $t+0$ plus the expiring
period based on ticket ids (e.g. ticket ID 5 allocated in slot 0 is expiring in slot “5 + expiring period”).
Another question is if tickets only start to expire once they are sold. In the simulation they start to expire
once they are on the market. Furthermore, the question arises if they are certain to be allocated at some
slot within the expiring period to reduce the complexity of the pricing. In our simulation this is not the
case as this would have second order effects on the allocation mechanism making it more sophisticated.
Hence, tickets on the market lose value with every slot as the probability of getting assigned lowers by
$(1/expiry\_period)$ every slot.

Furthermore, the pricing becomes more complicated for the bidders in case of expiring tickets as a
discount has to be considered for the ticket value. Given that the tickets have a static expiry period, for the
primary market the value can be statically calculated as the value of the ticket $V(ticket)$ times probability
of the ticket getting assigned to a slot $P(Assigned Slot)$. The expected value (EV) of a ticket can be
defined as:
\[ EV(ticket) = V(ticket) \ast P(Assigned~Slot) = V(ticket) \ast \left( 1 - \left( 1 - \frac{X}{Z}\right)^{\frac{S}{X}} \right) \]

With $X$ denoting the number of slot for tickets to be assigned to (e.g. 32 per epoch), $Z$ denoting the total
number of tickets in circulation (\textit{max\_tickets} in the implementation) and $S$ denoting the total slots a ticket
is valid (\textit{expiry\_period} in the implementation).

It is to note that the combination of expiring and reimbursable tickets seems not to be desirable, as this
would mean that the protocol needs to price for the expiry date, needing to assume a discount rate (cost of
capital), which could be complicated to properly define on protocol level. Hence, we have deemed such a
configuration not feasible.

\subsubsection{Price Volatility Effects}
\label{subsubsec:PriceVolatilityEffects}

In order to realistically model the bidding dynamics builder specialization has to be taken into account. As
shown by \citep{heimbach_non-atomic_2024} and \citep{gupta_centralizing_2023} certain builders drive in high CEX-DEX-price
volatility environments due to advantageous proprietary orderflow (e.g. integrated builder-searchers).
Further, \citep{oz_who_2024} have shown the rising importance of diverse order flow.

To avoid overcomplicating the volatility simulation a log-normal distribution of volatility for each slot is
assumed in line with existing literature (e.g. \citep{liu_volatility_2019,tegner_volatility_2018}). The
constant \textit{price\_vola} is used to set the values for it. If it is set to [\textit{None, None}] it is disabled. Otherwise the
first value represents the mean, the second the sigma of the distribution. Standard values that are used are
[0, 0.2]. It is to note that this averages slightly above 1 and hence \textit{sys\_params['expected\_vola']} is
calculated as the benchmark ticket holders use for their volatility calculations.

Further, for each builder a \textit{vola\_spec\_factor} representing the volatility specialization of the builder is set
with a random normal distribution with mean 1 and a standard deviation that can be adjusted (a value >1
indicates that the builder drives in high-volatility environments and vice versa). The adjustment is then
calculated as follows:
\[ Adj_{vola} = 1 + (Vola~per~slot - expected~vola) \ast vola~spec~factor \]

This results in a positive adjustment in high volatility slots and a negative adjustment in low volatility
environments.

\subsection{Simulation Results}
\label{subsec:SimulationResults}

\subsubsection{Results on first configuration “Simple FPA Auction”}
\label{subsubsec:ResultsOnFirstConfigurationSimpleFPAAuction}

Running the simulation setup with the first configuration “Simple FPA Auction” we can observe the
following results:

\begin{table}[H]
\centering
\renewcommand{\arraystretch}{1.25} % Row height adjustment
\setlength{\arrayrulewidth}{0.15mm} % Thick lines
\setlength{\tabcolsep}{0.5em}

% \caption{}
% \label{}

\adjustbox{max width = \linewidth}{
\begin{tabular}{|Z{4cm}|Z{5cm}|Z{2.25cm}|Z{2cm}|}

\hline
\multicolumn{4}{|l|}{\textbf{Simulation results Simple FPA Auction} }\\ [-1mm]
\multicolumn{4}{|l|}{ \textit{(from run 2024-09-17\_17-10 UTC, 10 runs, 1000 time steps)} }\\ \hline

\textbf{Objective} & \textbf{Metric} & \textbf{Results} & \textbf{Evaluation} \\ \hline

\multirow{3}{*}{Decentralization} & Market share & 93.2\% & \color{red} Low \\ \cline{2-4}
& Nakamoto-coefficient & 1 & \color{red} Low \\ \cline{2-4}
& Herfindahl-Hirschman Index & 8718.0 & \color{red} Low \\ \hline

MEV Capture & MEV-Share Protocol & 79.8\% / 89.1\% & \color{green!50!black} High \\ \hline
Price Predictability & GK Measure & 0.008 & \color{green!50!black} High \\ \hline
Price Smoothness & $V(\Delta p)$ & 0.026 & \color{green!50!black} High \\ \hline
Price Accuracy & MEV-Share Protocol & 79.8\% & \color{green!50!black} High \\ \hline

\end{tabular}}
\end{table}

We observe a strong centralization pressure on all relevant metrics. At the same time we see a high MEV
capture and desirable metrics on the price behavior. Based on the data we can conclude that small
differences in MEV extraction capabilities have strong centralization forces due to the fact that a period of
32 slots ahead results in a bidding behavior driven by expected block values. This is in line with previous
theoretical findings, e.g. \citep{bahrani_centralization_2024}. As shown in \citep{stichler_does_2024}, there is only a very limited
correlation of MEV a few slots ahead. The high MEV capturing ability derives from the fact how the
bidding algorithm is designed. As described above we have worked with a simple bidding of intrinsic
valuations. This might play out differently in reality considering that the winning ticket holders might
observe retrospectively the other bids and adjust his bidding accordingly. The conclusions are under the
assumption that the first price auction can be run privately, if the auction will be with public bids, it
essentially turns into a second price auction, as outlined above.
\begin{figure}[H]
    \centering
    \includegraphics[width=0.95\linewidth]{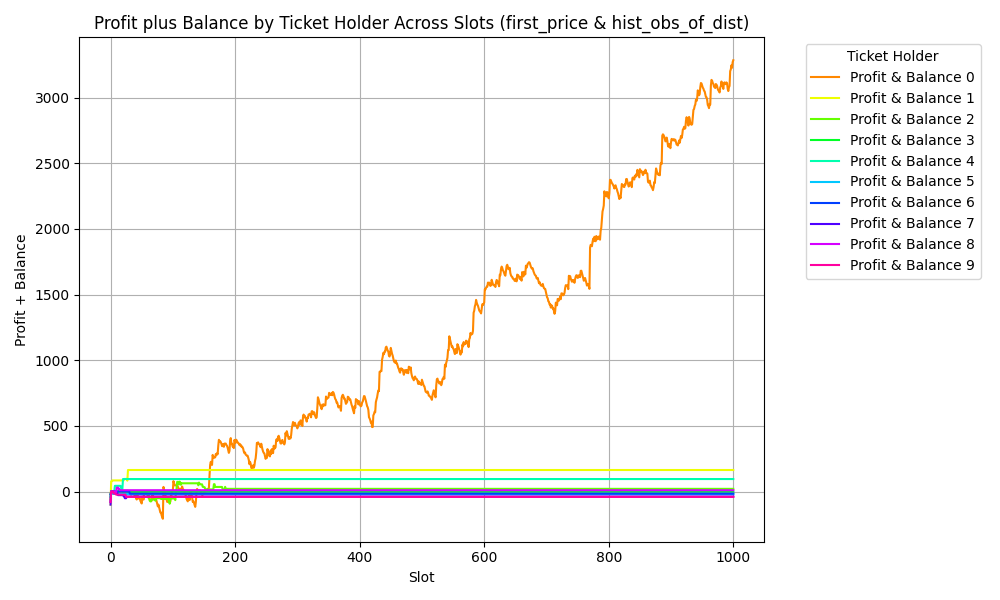}
    %\caption{}
    %\label{}
\end{figure}

Exemplary analysis of the profit plus balance per holder shows that one holder with the best MEV
capturing abilities (in this case holder 0), clearly exceeds all other bidders.

Further it needs to be noted that it might not be realistic to assume that a secondary market can be
prevented.

Running the simulation with a secondary market enabled we can observe that the trends remain similar,
however the centralization is less strong with e.g. an average highest market share of 64\%\footnote{\textit{Run 2024-09-24\_07-48 UTC, 10 runs, 1000 time steps}}. This is
caused by the just-in-time secondary market auction, where specialized ticket holders purchase tickets in
low-/high volatility environments. Furthermore, from a theoretical perspective MEV capture should
increase as the secondary market opens optionality for the ticket holder and thereby increases the value.
This, however, is not included in the simulation for brevity reasons (in the quoted simulation MEV
capture is at 80\%). More details can be found in the simulation results in the code basis.

\clearpage
\subsubsection{Results on second configuration “JIT Second Price Slot Auction”}
\label{subsubsec:ResultsOnSecondConfigurationJITSecondPriceSlotAuction}

Running the second configuration we can observe the following results:
\begin{table}[H]
\centering
\renewcommand{\arraystretch}{1.25} % Row height adjustment
\setlength{\arrayrulewidth}{0.15mm} % Thick lines
\setlength{\tabcolsep}{0.5em}

% \caption{}
% \label{}

\adjustbox{max width = \linewidth}{
\begin{tabular}{|Z{4cm}|Z{5cm}|Z{2.25cm}|Z{2cm}|}

\hline
\multicolumn{4}{|l|}{\textbf{Simulation results JIT Second Price Slot Auction} } \\ [-1mm]
\multicolumn{4}{|l|}{\textit{(from run 2024-09-24\_10-52 UTC, 10 runs, 1000 time steps)} } \\ \hline

\textbf{Objective} & \textbf{Metric} & \textbf{Results} & \textbf{Evaluation} \\ \hline

\multirow{3}{*}{Decentralization} & Market share & 52.2\% & \color{orange} Medium \\ \cline{2-4}
& Nakamoto-coefficient & 1 & \color{red} Low \\ \cline{2-4}
& Herfindahl-Hirschman Index & 4741.7 & \color{orange} Medium \\ \hline

MEV Capture & MEV-Share Protocol & 76.6\% / 77.5\% & \color{green!50!black} High \\ \hline
Price Predictability & GK Measure & 3.8 & \color{red} Low \\ \hline
Price Smoothness & $V(\Delta p)$ & 1398.6 & \color{red} Low \\ \hline
Price Accuracy & MEV-Share Protocol & 77.5\% & \color{green!50!black} High \\ \hline

\end{tabular}}
\end{table}

Running the second configuration, which most closely represents the current MEV-Boost market, we can
observe several trends. The decentralization is inline with the currently observed decentralization in the
block builder market. As shown in the images below, the market shares in the simulation are very similar
to the builder market shares of the last 14 days\footnote{Image taken from mevboost.pics on 2024/09/24}. We observe one dominant player with one secondary
strong player. MEV capture is medium to high, due to the missing competition in the second price auction
some MEV is left on the table. The price attributes (predictability and smoothness) are low as the prices
fluctuate in accordance with the currently available MEV. Given the JIT modality price accuracy should
be high, however the missing competition in the second-price auction reduces the accuracy.

%\clearpage

\begin{figure}[H]
    \centering
    \includegraphics[width=0.98\linewidth]{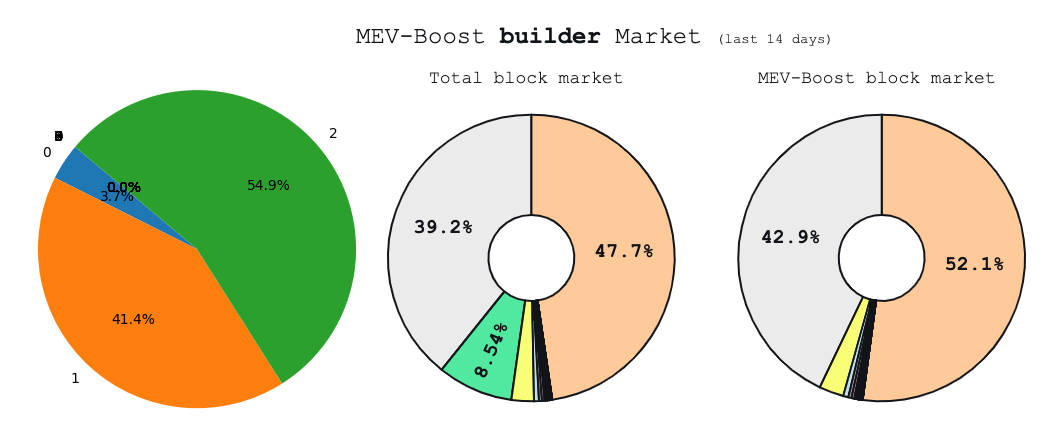}
    %\caption{}
    %\label{}
\end{figure}

\clearpage
\subsubsection{Results on third configuration “Flexible 1559-Style”}
\label{subsubsec:ResultsOnThirdConfigurationFlexible1559-Style}

Running the third configuration we can observe the following pattern.
\begin{table}[H]
\centering
\renewcommand{\arraystretch}{1.25} % Row height adjustment
\setlength{\arrayrulewidth}{0.15mm} % Thick lines
\setlength{\tabcolsep}{0.5em}

% \caption{}
% \label{}

\adjustbox{max width = \linewidth}{
\begin{tabular}{|Z{4cm}|Z{5cm}|Z{2.25cm}|Z{2cm}|}

\hline
\multicolumn{4}{|l|}{\textbf{Simulation results Flexible 1559-Style} } \\ [-1mm]
\multicolumn{4}{|l|}{\textit{(from run 2024-09-24\_19-07 UTC, 10 runs, 1000 time steps\tablefootnote{With EIP-1559\_max\_tickets = 4 and EIP-1559\_adjust\_factor = 8})} } \\ \hline

\textbf{Objective} & \textbf{Metric} & \textbf{Results} & \textbf{Evaluation} \\ \hline

\multirow{3}{*}{Decentralization} & Market share & 76.2\% & \color{red} Low \\ \cline{2-4}
& Nakamoto-coefficient & 1 & \color{red} Low \\ \cline{2-4}
& Herfindahl-Hirschman Index & 6062.7 & \color{red} Low \\ \hline

MEV Capture & MEV-Share Protocol & 40.2\% / n/a\% & \color{red} Low \\ \hline
Price Predictability & GK Measure & 1.095 & \color{orange} Medium \\ \hline
Price Smoothness & $V(\Delta p)$ & 91602.8 & \color{red} Low \\ \hline
Price Accuracy & MEV-Share Protocol & 40.2\% & \color{red} Low \\ \hline

\end{tabular}}
\end{table}

Running an 1559-style simulation several aspects become apparent. Firstly, it shows that a 1559-style
pricing is prone to oscillating price adaptation effects. Driven by the fact that several tickets can be sold
but only one gets taken off the market as well as the stepwise price adaptation, price swings become
self-reinforcing. This indicates that a 1559-style pricing needs to be very carefully designed and in our
simulation it was very hard to set the parameters ($EIP-1559\_max\_tickets$ \& $EIP-1559\_adjust\_factor$)
robustly. On the objectives and success metrics we observe a high level of centralization. Even though on
the primary market more ticket holders are able to purchase tickets in low-price periods (assuming a
random latency), the tickets are then sold in the secondary market. So, earnings are more decentralized
however the block builder is still highly centralized. With regards to price predictability, smoothness and
accuracy we also observe low the at best medium values, driven by the static price adaptation mechanism.

Preliminary, we can conclude that based on the simulation a 1559-style pricing is not an ideal pricing
mechanism for Execution Tickets.

\subsubsection{Results on fourth configuration “Fixed SPA”}
\label{subsubsec:ResultsOnFourthConfigurationFixedSPA}

Running the fourth configuration we can observe the following metrics.
\begin{table}[H]
\centering
\renewcommand{\arraystretch}{1.25} % Row height adjustment
\setlength{\arrayrulewidth}{0.15mm} % Thick lines
\setlength{\tabcolsep}{0.5em}

% \caption{}
% \label{}

\adjustbox{max width = \linewidth}{
\begin{tabular}{|Z{4cm}|Z{5cm}|Z{2.25cm}|Z{2cm}|}

\hline
\multicolumn{4}{|l|}{\textbf{Simulation results Fixed Second Price Auction}} \\ [-1mm]
\multicolumn{4}{|l|}{\textit{(from run 2024-09-25\_08-43 UTC, 10 runs, 1000 time steps) }} \\ \hline

\textbf{Objective} & \textbf{Metric} & \textbf{Results} & \textbf{Evaluation} \\ \hline

\multirow{3}{*}{Decentralization} & Market share & 69.8\% & \color{red} Low \\ \cline{2-4}
& Nakamoto-coefficient & 1 & \color{red} Low \\ \cline{2-4}
& Herfindahl-Hirschman Index & 5950.9 & \color{red} Low \\ \hline

MEV Capture & MEV-Share Protocol & 69.8\% / 81.7\% & \color{orange} Medium  \\ \hline
Price Predictability & GK Measure & 0.006 & \color{green!50!black} High \\ \hline
Price Smoothness & $V(\Delta p)$ & 0.012 & \color{green!50!black} High \\ \hline
Price Accuracy & MEV-Share Protocol & 69.8\% & \color{orange} Medium  \\ \hline

\end{tabular}}
\end{table}

Running this configuration we can observe low scores on decentralization and medium to high scores on
MEV capture and price attributes. The high concentration is driven by the auction being based on
expected values and no option to sell the tickets in a secondary market in this configuration. This leads to
the market participant with the most favorable MEV capturing abilities as the market leader. MEV capture
is average with the auction earnings being defined by the expected valuation of the second highest bidder.
Further, a discount accrues due to the market only being based on expected valuations for future MEV
revenues. This however, makes the price very predictable and stable.

Overall, we don't see this as an ideal configuration, as it leads to high centralization and at the same time
prohibiting a secondary market might not be practically feasible.

\subsubsection{Results on fifth configuration “Flexible, refundable AMM”}
\label{subsubsec:ResultsOnFifthConfigurationFlexibleRefundableAMM}

\begin{table}[H]
\centering
\renewcommand{\arraystretch}{1.25} % Row height adjustment
\setlength{\arrayrulewidth}{0.15mm} % Thick lines
\setlength{\tabcolsep}{0.5em}

% \caption{}
% \label{}

\adjustbox{max width = \linewidth}{
\begin{tabular}{|Z{4cm}|Z{5cm}|Z{2.25cm}|Z{2cm}|}

\hline
\multicolumn{4}{|l|}{\textbf{Simulation Results Flexible, Refundable AMM} }\\ [-1mm]
\multicolumn{4}{|l|}{ \textit{(from run 2024-09-25\_15-10 UTC, 10 runs, 1000 time steps, AMM\_adjust\_factor: 25) }}\\ \hline

\textbf{Objective} & \textbf{Metric} & \textbf{Results} & \textbf{Evaluation} \\ \hline

\multirow{3}{*}{Decentralization} & Market share & 89.0\% & \color{red} Low \\ \cline{2-4}
& Nakamoto-coefficient & 1 & \color{red} Low \\ \cline{2-4}
& Herfindahl-Hirschman Index & 7991.0 & \color{red} Low \\ \hline

MEV Capture & MEV-Share Protocol & 75.8\% / 84.0\% & \color{green!50!black} High  \\ \hline
Price Predictability & GK Measure & 0.010 & \color{green!50!black} High \\ \hline
Price Smoothness & $V(\Delta p)$ & 0.057 & \color{green!50!black} High \\ \hline
Price Accuracy & MEV-Share Protocol & 75.8\% & \color{green!50!black} High  \\ \hline

\end{tabular}}
\end{table}

In this configuration we can observe low scores on decentralization and medium to high scores on MEV
capture and price attributes. Similarly to previous configurations the intrinsic valuation of tickets leads to
one dominant player capturing the majority of the market and the non-existing secondary market does not
allow for specialized ticket holders to buy tickets in certain environments. The AMM-style pricing seems
to be able to settle in at a price that captures a medium to high level of MEV available and remains stable.

\subsubsection{Results on sixth configuration “Fixed, resellable FPA”}
\label{subsubsec:ResultsOnSixthConfigurationFixedResellableFPA}

Running the sixth configuration with a flexible AMM-style pricing with no expiry period and a secondary
market enabled and non-reimbursable. We can observe the following:

\begin{table}[H]
\centering
\renewcommand{\arraystretch}{1.25} % Row height adjustment
\setlength{\arrayrulewidth}{0.15mm} % Thick lines
\setlength{\tabcolsep}{0.5em}

% \caption{}
% \label{}

\adjustbox{max width = \linewidth}{
\begin{tabular}{|Z{4cm}|Z{5cm}|Z{2.25cm}|Z{2cm}|}

\hline
\multicolumn{4}{|l|}{\textbf{Simulation results Fixed, resellable FPA}} \\ [-1mm]
\multicolumn{4}{|l|}{\textit{(from run 2024-09-26\_08-18 UTC, 10 runs, 1000 time steps) }}\\ \hline

\textbf{Objective} & \textbf{Metric} & \textbf{Results} & \textbf{Evaluation} \\ \hline

\multirow{3}{*}{Decentralization} & Market share & 48.9\% & \color{orange} Medium \\ \cline{2-4}
& Nakamoto-coefficient & 1.9\tablefootnote{Note, that it is the average over 10 runs.} & \color{orange} Medium \\ \cline{2-4}
& Herfindahl-Hirschman Index & 3458.1 & \color{orange} Medium \\ \hline

MEV Capture & MEV-Share Protocol & 77.2\% & \color{green!50!black} High  \\ \hline
Price Predictability & GK Measure & 0.042 & \color{green!50!black} High \\ \hline
Price Smoothness & $V(\Delta p)$ & 0.521 & \color{green!50!black} High \\ \hline
Price Accuracy & MEV-Share Protocol & 77.2\% & \color{green!50!black} High  \\ \hline

\end{tabular}}
\end{table}

In this configuration we can observe that the decentralization metrics are on a medium level while the
MEV capture and price attributes are medium to high. Even though on the primary ticket market a
singular ticket buyer is usually successful on the secondary market a more diverse set of specialized ticket
holders buy the tickets and get the slots (as can be seen exemplary in the image below). This leads to slots
being more equally distributed however one actor earning the majority of rewards. The price seems
generally stable and predictable with an overall higher level of MEV capture than other configurations.

\begin{figure}[H]
    \centering
    \includegraphics[width=0.9\linewidth]{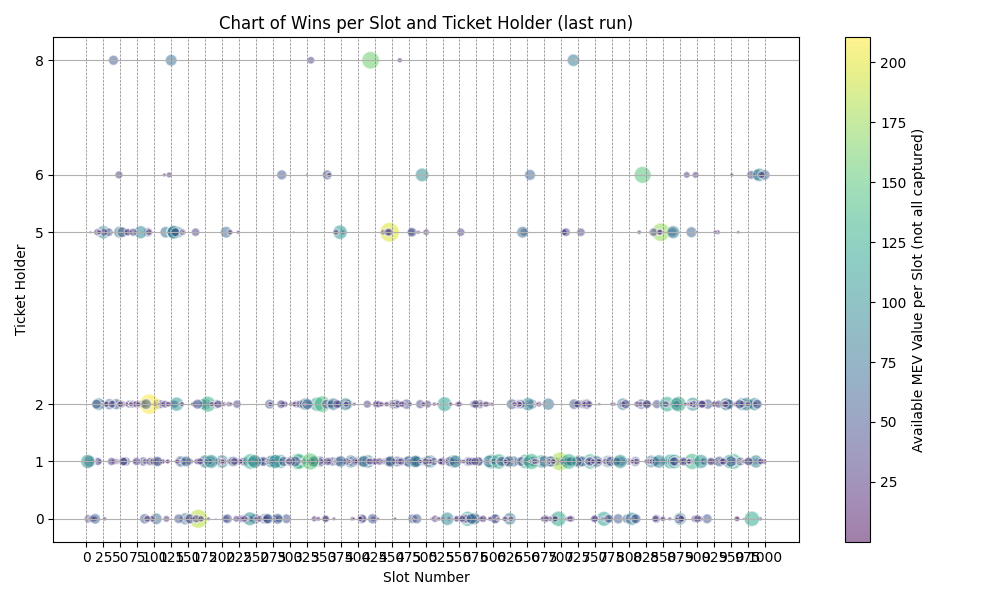}
    %\caption{}
    %\label{}
\end{figure}

\subsubsection{Summary on simulating the configurations}
\label{subsubsec:SummaryOnSimulatingTheConfigurations}

Comparing all configurations we can observe several trends.

\begin{table}[H]
\centering
\renewcommand{\arraystretch}{1.25} % Row height adjustment
\setlength{\arrayrulewidth}{0.15mm} % Thick lines
\setlength{\tabcolsep}{0.5em}

% \caption{}
% \label{}

\adjustbox{max width = \linewidth}{
\begin{tabular}{|Z{3cm}|Z{3cm}|M{2.25cm}|M{2.25cm}|M{2.25cm}|M{2.25cm}|M{2.25cm}|M{2.25cm}|}

\hline
\multicolumn{8}{|l|}{\textbf{Overview of Simulation Results}\tablefootnote{Color coding of results based on literature and subjective judgment}} \\ \hline

\textbf{Objective} & \textbf{Metric} & \textbf{Simple FPA auction} & \textbf{JIT SPA slot auction} & \textbf{Flexible 1559-style\%} & \textbf{Fixed SPA} & \textbf{Flexible, refundable AMM} & \textbf{Fixed, refundable FPA} \\ \hline

\multirow{3}{*}{Decentralization} & Market share & \color{red} 93.2\% & \color{orange} 52.2\% & \color{red} 76.2\% & \color{red} 69.8\% & \color{red} 89.0\% & \color{orange} 48.9\% \\ \cline{2-8}
 & Nakamoto-coefficient & \color{red} 1 & \color{red} 1 & \color{red} 1 & \color{red} 1 & \color{red} 1 & \color{orange} 1.9 \\ \cline{2-8}
 & Herfindahl-Hirschman Index & \color{red} 8718.0 & \color{orange} 4741.7 & \color{red} 6062.7 & \color{red} 5950.9 & \color{red} 7991.0 & \color{orange} 3458.1 \\ \hline
MEV Capture & MEV-Share Protocol & \color{green!50!black} 79.8\% / 89.1\% & \color{green!50!black} 76.6\% / 77.5\% & \color{red} 40.2\% & \color{orange} 69.8\% / 81.7\% & \color{green!50!black} 75.8\%  / 84.0\%  & \color{green!50!black} 77.2\%  \\ \hline
Price Predictability & GK Measure & \color{green!50!black} 0.008 & \color{red} 3.8 & \color{orange} 1.095 & \color{green!50!black} 0.006 & \color{green!50!black} 0.010 & \color{green!50!black} 0.042 \\ \hline
Price Stinosc & $V(\Delta p)$ & \color{green!50!black} 0.026 & \color{red} 1398.6 & \color{red} 91602.8 & \color{green!50!black} 0.012 & \color{green!50!black} 0.057 & \color{green!50!black} 0.521 \\ \hline
Price Accuracy & MEV-Share Protocol & \color{green!50!black} 79.8\% & \color{green!50!black} 77.5\% & \color{red} 40.2\% & \color{orange} 69.8\% & \color{green!50!black} 75.8\% & \color{green!50!black} 77.2\% \\ \hline

\end{tabular}}
\end{table}

It generally shows that in all configurations decentralization remains a challenge. None of the
configuration scores particularly well on the decentralization metrics. This is driven by the diverse
abilities of ticket holders and the fact that in most scenarios the bids are based on expected valuation
which leaves out specialization factors. Here it shows that in cases with a secondary market enabled the
centralization is reduced as in JIT auctions specialized ticket holders buy the tickets for the same slot.

\clearpage

With regards to MEV capture we can see different attributes emerge. The auction formats generally score
well, similarly the AMM-style pricing scores well. The 1559-style pricing is capturing less MEV due to a
step-wise and less dynamic price adaptation mechanism.

With regards to the price predictability, smoothness and accuracy we can observe that the auction formats
that operate with a longer lookahead are very predictable and smooth, while JIT auctions and a 1559-style
pricing are less smooth. The 1559-style pricing is a specialized case more in depth discussed below.

\subsection{Evaluation of Parameters}
\label{subsec:EvaluationOfParameters}

In this chapter we will evaluate each parameter and its possible values and what we have observed and
concluded from the simulation on it. It is to note that on the target amount of tickets and enhanced
lookahead, we did not run this realistically in the simulation and hence rely on the theoretical evaluation
presented above on this.

\subsubsection{Findings on Auction Formats}
\label{subsubsec:FindingsOnAuctionFormats}

\paragraph{First Price Auctions}
\label{par:FirstPriceAuctions}

With regards to first price auctions, we saw a “winner's curse” play out, in the terms that assuming that
bidders have differing intrinsic value expectations for a ticket which are following a normal distribution,
the most optimistic bidders wins. And the most optimistic bidder with the highest valuation overestimates
the value the most and thereby makes a loss on the trade. This is a known problem of auctions (e.g.
\citep{bazerman_i_1983}). However, noteworthy to point out, as this leads to higher
“risk-adjustments” by bidders which in turn could lead to reduced MEV capture by the protocol.

With regards to the simulation it shows that first price auctions generally perform well, however two
things need to be critically challenged here. Firstly, in order to successfully run a first price auction it
needs to be done with sealed bids. As outlined above, several proposals for this are currently being
discussed, however still in the earlier stages. Secondly, in the simulation the bidding is based on a bidding
of intrinsic valuations for a first price auction with no further information. This assumption will not hold
true in multi-round scenarios of execution ticket selling. So more sophisticated bidding strategies based
on historical bids of competitors might emerge that might potentially reduce the captured MEV. So it is
unclear yet, if first price auctions can be actually designed in this scenario in a way to behave differently
than second price auctions.

\paragraph{Second Price Auctions}
\label{par:SecondPriceAuctions}

For second price auctions we observed that the MEV capture highly depends on the competitiveness of
the specific simulation. In cases with at least two similarly strong ticket holders the MEV capture was
high, however on average it was only medium, given the missing competition.

\paragraph{EIP-1559 style pricing}
\label{par:EIP1559StylePricing3}

As outlined above the EIP-1559 pricing needs to be adapted to suit for Execution Tickets and we have
implemented it as a batch process. However, we observe that this leads to self-reinforcing oscillating
ticket prices. Even adjusting the \textit{EIP-1559\_adjust\_factor} does not lead to better outcomes in our
simulations. This leads to the conclusion that a batch update process is not sufficient. How a continuous
price update process can be technically implemented in a decentralized setting remains an open question.
Overall, the pricing mechanism needs to be carefully designed to achieve a desired price behavior.

\clearpage
\begin{figure}[H]
    \centering
    \includegraphics[width=0.85\linewidth]{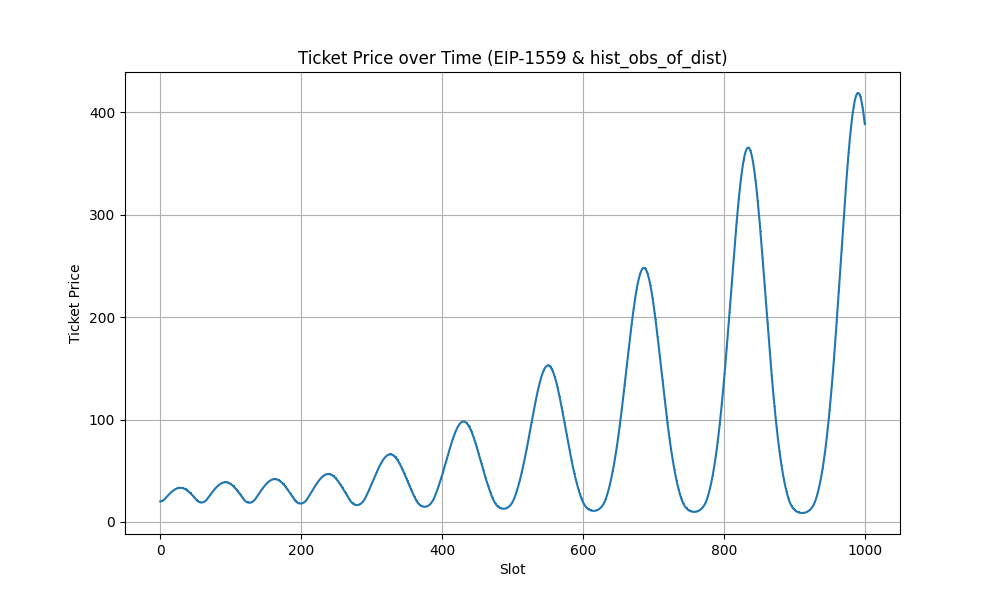}
    %\caption{}
    %\label{}
\end{figure}

Further, we observe that more than for other pricing mechanisms the initial ticket price is crucial, as a too
low or too high initial price leads to a long price finding period.

Further, in certain simulations\footnote{E.g. see simulation results 2024-05-14\_18-09\_1\_1000\_EIP-1559 for details} we have observed that if one ticket holder has a significantly higher willingness to pay than others, the prices stabilize at a point where only this ticket holder is able to purchase tickets, leading to a high centralization.

\paragraph{AMM-style Pricing}
\label{par:AMMStylePricing}

For the AMM-style pricing as outlined above it needs to be adapted to be suitable for Execution Tickets.
As described above we have implemented the “Continuous e delta update function”. It becomes apparent
that the selection of the constant b and the adjustment steps of the AMM (\textit{AMM\_adjust\_factor}) need to be
carefully designed. Running configurations with AMM–style pricing shows that the pricing mechanism
can be sensitive to the adjustment factor. A too slow adoption does not accurately capture the demand, a
too large adaption factor is not granular enough to differentiate the expected valuations and would lead to
a latency race.

Running the AMM-style simulation with reimbursement (20\% discount) it can be observed that even
before the first ticket is redeemed the long tail builders give back all their tickets to the protocol, which
are then bought by the top builders. This underlines that for an AMM-style pricing the initial price needs
to be carefully chosen.

However, the simulations show promising results that this mechanism is able to capture a high level of
MEV. From an operational perspective it remains to be investigated how this could be implemented to suit
the selling process needed for Execution Tickets.

\paragraph{Conclusion on auction formats}
\label{par:ConclusionOnAuctionFormats}

Taking into consideration the different observations, based on the simulation results we conclude that an
auction based format, most probably second price auction, is the most feasible format. It leads to a high
captured MEV, is DSIC and leads to favorable price properties in the simulation. A first price auction also
shows promising results in the simulation on captured MEV, however is not DSIC. An AMM-style
pricing seems also to be a promising solution, however more open design mechanism and implementation
questions remain.

One relevant question remains around OCA-proofness, in case the ET earnings are burned. There might
be a sybil attack vector where block builders bribe the actor / committee defining the winning bid and
thereby being able to achieve a lower price. E.g. if the winning bid is 10 ETH, the block builder however
pays the committee members 5 ETH to artificially set the winning bid price at 1 ETH, there could be a 4
ETH profit margin. To avoid this bids or prices would need to be on-chain which is not feasible given the
time horizon. Another option could be a leaderless auction as outlined by \citep{white_leaderless_2024}.

\subsubsection{Findings on Amount of Tickets (fixed vs. flexible)}
\label{subsubsec:FindingsOnAmountOfTicketsFixedVsFlexible}

We observe that this attribute is closely related to the pricing mechanism. For certain mechanisms fixed
amounts of tickets make more sense (auctions) while for others (EIP-1559 and AMM-style) a flexible
amount makes more sense. Hence, we see this more as a secondary attribute that is deducted from the
pricing mechanism.

\subsubsection{Findings on Expiring Tickets}
\label{subsubsec:FindingsOnExpiringTickets}

In the simulation we observe that especially for short expiry times the MEV capture is impaired, as ticket
buyers need to discount the value of a ticket on the primary market and on the secondary market since the
possibility of a ticket expiry without redemption needs to be priced in. This leads to generally lower
captured MEV values.

Further, we observe that it has secondary complications as the pricing of each actor becomes more
sophisticated as the expiry period, outstanding tickets etc. need to be factored in. This leads to the
conclusion that non-expiring tickets seem to be the favorable configuration.

\subsubsection{Findings on Refundability}
\label{subsubsec:FindingsOnRefundability}

Regarding refundability we only observe limited effects on the market dynamics with the tested discount
(around 20\%). It leads to more security for ticket holders, however this depends on the discount. Further
it is closely related with the secondary market, in case a secondary market exists this option is often more
attractive to dispose of tickets.

It shows that allowing for refundability does not influence the mechanism in a substantial way and
complicates the design choices as well as the decisions for ticket holders. Hence, the preliminary analysis
leads to the conclusion that tickets shall not be refundable.

\subsubsection{Findings on Resalability (Secondary Market)}
\label{subsubsec:FindingsOnResalabilitySecondaryMarket}

Regarding the secondary market an interesting finding is that this increases decentralization. Due to the
ability of more specialized ticket holders to buy tickets just-in-time in periods where they are able to
capture higher MEV due to specialization.

Further, it leads to overall higher MEV captured due to reduced risks for the primary ticket holders. This
is however not modeled in depth in the simulation.

Additionally, we observe that in some configurations with discrete pricing (e.g. AMM-style pricing) it
leads to arbitrage opportunities, if the AMM-pricing is not adapting fine granularly enough and tickets
can be bought by the ticket holder with the lowest latency and then be resold at a higher price at the
secondary market.

Given that also from a technical perspective it is difficult to prevent a secondary market, a preliminary
recommendation is rather embrace the benefits of it and try to foster it.

\clearpage
\subsection{Further Simulation Observations}
\label{subsec:FurtherSimulationObservations}

\subsubsection{Findings on Decentralization}
\label{subsubsec:FindingsOnDecentralization}

Considering the fact that Execution Tickets are auctioned off ahead of time to avoid the valuation in flight
issue, it leads to the bidder(s) with the highest average expected valuation to dominate the market. This
leads to a high concentration of tickets in the hands of a few ticket holders. If a functioning secondary
market is enabled, this reduces the effect at the actual block building level by a more diverse set of
specialized ticket holders buying tickets on the secondary market. It nevertheless leaves several avenues
for multi-block MEV and/or censorship open. The high centralization forces are in line with the findings
of \citep{bahrani_centralization_2024} showing that the market decentralization highly depends on the MEV extraction
capabilities of the top builders.

\subsubsection{Impact of Ticket Holder attributes}
\label{subsubsec:ImpactOfTicketHolderAttributes}

Generally we observe that the attributes of the ticket holders highly influence the outcome of the
simulation. Even small adjustments in the MEV capturing abilities, aggressiveness and/or volatility
specialization lead to very different outcomes. This highlights again that a competitive builder market is
important to avoid monopolization.

%% file: Sections/06-MitigationOfMultiBlockMEV.tex
\section{Mitigation of Multi-Block MEV}
\label{sec:MitigationOfMultiBlockMEV}

As introduced in Chapter \ref{sec:IntroductionMotivation} multi-block MEV is one of the major concerns in the context of Execution
Tickets. With the acceptance of a certain level of centralization amongst execution block proposers the
occurrence for multi-slot assignment will increase and especially in configurations with secondary
markets the ability to purchase consecutive slots is high. It is however to note, that already at the current
state several consecutive slots per block proposer are frequent and no extreme MMEV capturing
strategies have been observed yet \citep{stichler_does_2024}.

Nevertheless, to combat this, several angles outlined below seem feasible.

\subsection{Missed Block Penalties}
\label{subsec:MissedBlockPenalties}

To avoid a block proposer purchasing several slots in a row and leaving them empty to capture all MEV
in the end, it might be feasible to introduce missed slot penalties that exponentially increase over
consecutive slots missed. This poses the question of how the penalty can be enforced. As also
enforcement mechanisms are needed for safety violations, it might result in the necessity of introducing a
staking requirement for execution ticket holders.

\subsection{Inclusion Lists, FOCIL and AUCIL}
\label{subsec:InclusionListsFOCILandAUCIL}

Inclusion Lists (IL) might be a feasible way to combat this issue. They might not only be a good tool for
censorship resistance, but also to prevent multi-block MEV. The Ethereum community is currently
discussing to enshrine inclusion lists (EIP-7547) in the protocol\footnote{\url{https://eips.ethereum.org/EIPS/eip-7547} retrieved on 03/03/2024 \newline \url{https://ethresear.ch/t/inclusion-list-eip-7547-end-to-end-workflow/18810}, \newline \url{https://ethresear.ch/t/no-free-lunch-a-new-inclusion-list-design/16389} \& \newline \url{https://gist.github.com/michaelneuder/ba32e608c75d48719a7ecba29ec3d64b} retrieved on 26/09/2024}. The inclusion is envisioned as a
forward inclusion list, where the proposer of slot N can specify a list that needs to be included either in
slot $N$, slot $N+1$ or slot $N+32$. The inclusion list needs to be submitted at time of the block proposal, to
avoid exploiting MEV opportunities arising between their slot and the inclusion list slot. More
sophisticated versions of inclusion lists have been recently proposed with FOCIL and AUCIL with
stronger censorship-resistance properties.\footnote{Note that currently the maximum number of transactions per inclusion list is proposed at 16} \footnote{\url{https://ethresear.ch/t/fork-choice-enforced-inclusion-lists-focil-a-simple-committee-based-inclusion-list-proposal/19870} retrieved on 26/09/2024} \footnote{\url{https://ethresear.ch/t/aucil-an-auction-based-inclusion-list-design-for-enhanced-censorship-resistance-on-ethereum/20422} retrieved on 26/09/2024}

With this setup empty blocks can be avoided and the proposer in the respective slot must at least include
the inclusion list transactions. In an Execution Ticket mechanism design, the functioning of ILs need to be
revised. As \citep{drake_session_2023} proposed, they can then be proposed by the beacon chain validator and
forwarded to the execution chain validator. If the execution chain validator needs to include them in slot $N$
and/or $N+1$ or $N+32$ is still subject of debate. This general concept however is crucial, as it breaks the
monopoly of a proposer holding tickets over several consecutive slots on the included transactions. Given
for example, a malicious execution chain proposer is holding back Uniswap transactions for a certain
token pair to deviate the price on Uniswap from the price on a centralized exchange and exploit this price
difference in the last slot, the proposer always has to fear that via an enforced IL transaction is snatched
away. This increases the uncertainty and prevents the malicious proposer from investing manipulation
capital in the first place. It is to note, however, that this greatly depends on the specific implementation of
inclusion lists. Assuming for example that the execution chain proposers only need to include transactions
in slot $N+1$ at a non-specified block position, they can monitor IL transactions and capture built-up MEV
before. For the sake of this document, it is assumed that an IL mechanism can be scoped that prevents
such loopholes by effective mechanism design and by setting effective penalties.

Summarizing, it can be observed that missed penalty slots in combination with inclusion lists deem to be
a potential solution to avoid multi-slot MEV, as it introduces sufficient uncertainty for execution block
proposers that prevents them from investing manipulation capital into multi-slot MEV strategies.\footnote{It is to note that a Prevention Paradox might occur here. By inserting a multi-slot MEV preventive transaction, the
multi-slot MEV strategy is successfully prevented, which however cannot be shown as it is non-existent.}

\subsection{Concurrent Block Proposers (Multiplicity)}
\label{subsec:ConcurrentBlockProposersMultiplicity}

Another possibility to reduce the risk for multi-slot MEV is the introduction of multiple concurrent block
proposers (multiplicity). Recent discussions have emerged around having multiple concurrent block
proposers for one slot to increase censorship resistance \citep{neuder_concurrent_2024,resnick_braid_2024}. This
would also translate into multi-slot MEV resistance as an actor aiming to extract multi-slot MEV needs to
control concurrent slots in order to ensure that no one else is taking advantage of the MEV opportunity.
As it seems unlikely to be implemented soon and the compatibility with Execution Tickets is unclear, we
will not investigate this option in-depth in this context.

%% file: Sections/07-Conclusion.tex
\section{Conclusion}
\label{sec:Conclusion}

Execution Tickets present a promising pathway for enhancing the Ethereum block space allocation
mechanism. With the appropriate mechanism design, they have the potential to foster decentralization
among beacon chain validators and enable MEV capture at the protocol level. Furthermore, they can
promote reward smoothing while simultaneously reducing the risk of MEV-stealing.

From a game-theoretical perspective, it is crucial to select a mechanism that satisfies the criteria Block
Producer Incentive Compatibility (BPIC), Dominant Strategy Incentive Compatibility (DSIC), and
Off-Chain Agreement proofness (OCA-proofness). The mechanism should maximize protocol-level MEV
capture while fostering decentralization among beacon chain validators and to a certain extent execution
chain block builders.

Our theoretical framework identifies three key areas of relevant objectives: Optimization Parameters,
Operational Processes, and Pricing Behavior, each encompassing several sub-objectives. The primary
optimization parameters are decentralization, MEV capture, and Block Producer Incentive Compatibility
(BPIC). We propose measuring decentralization using the metrics highest market share of a single ticket
holder, Nakamoto coefficient, and Herfindahl-Hirschman Index, while MEV capture is measured by the
protocol's MEV share of ticket holder rewards. BPIC requires theoretical evaluation.

Regarding the operational process, several relevant parameters need qualitative evaluation. For pricing
behavior, the relevant objectives are price predictability, smoothness, and accuracy. These can be
measured using the Garman-Klass (GK) Measure, variance of price changes (deltas), and the protocol's
MEV share.

In terms of mechanism design, we identify seven primary parameter choices: amount of tickets, expiry of
tickets, refundability, resalability, enhanced lookahead, pricing mechanism, and target amount of tickets.
We outlined the design space of potential parameter values and proposed four different pricing
mechanisms resulting in 640 potential mechanism configurations. Based on these, we propose six
concrete promising mechanism configurations.

We analyzed the mechanisms both theoretically and through an agent-based simulation, conducting over
300 simulation runs. From this analysis, we conclude several findings. None of the mechanisms scored
particularly well on decentralization due to the varying capabilities of ticket holders and reliance on
expected valuations for bids. However, enabling a secondary market reduces centralization, as specialized
ticket holders purchase tickets for specific slots in just-in-time (JIT) auctions. 

Regarding MEV capture, auction formats and AMM-style pricing generally perform well. In contrast, the
EIP-1559-style pricing captures less MEV due to its less dynamic price adaptation mechanism. In terms
of price predictability, smoothness, and accuracy, auction formats with longer lookahead periods are very
predictable and smooth. Conversely, JIT auctions and EIP-1559-style pricing exhibit less smoothness.
Based on these findings, we conclude that a second-price auction format is the most feasible mechanism.
First-price auctions show promising results in the simulation as well, however leave more questions open
from a theoretical mechanism perspective. AMM-style pricing mechanism also appears promising;
however, several design and implementation questions remain open, requiring further research.

We observe that the choice between a fixed or variable number of tickets is closely linked to the pricing
mechanism. Certain mechanisms, such as auctions, are better suited to a fixed number of tickets, while
others, like EIP-1559 and AMM-style pricing, may benefit from a flexible ticket amount.

Our simulation indicates that short expiry times impair MEV capture, as ticket buyers discount the value
due to the risk of expiration without redemption. This leads to more complex pricing strategies and
generally lower MEV capture. Therefore, we conclude that non-expiring tickets are the favorable
configuration.

Regarding refundability, we observed limited effects on market dynamics with the tested discount of
around 20\%. While refundability provides more security for ticket holders, its impact depends on the
discount rate. Consequently, we recommend that tickets should not be refundable to avoid complicating
the mechanism design.

\clearpage

Enabling a secondary market increases decentralization by allowing specialized ticket holders to purchase
tickets just-in-time, reducing risks for primary ticket holders and thereby leading to higher MEV capture.
Nevertheless, in line with \citep{bahrani_centralization_2024} we observe that the decentralization of the builder market
highly depends on the MEV extraction capabilities of the top builders. Given the technical difficulty of
preventing a secondary market, we preliminarily recommend embracing and fostering its benefits.

A primary concern with Execution Tickets is the risk of multi-slot MEV. Historically, no large-scale
multi-slot MEV strategies have been observed. Our preliminary evaluation suggests that implementing a
combination of missed slot penalties and inclusion lists could feasibly reduce this risk. However, further
research is required to specify the exact mechanisms for these mitigations.

Overall, Execution Tickets deem to be a feasible mechanism to sell block space. Concerns around
execution ticket holder decentralization, OCA-proofness and multi-block MEV remain. Auction formats
or an AMM-style pricing mechanism show initial promising results, however the pricing mechanism
needs further careful consideration.

%% file: Sections/A-NumberOfExecutionTickets.tex
\section*{A. Number of Execution Tickets}
\addcontentsline{toc}{subsection}{A. Number of Execution Tickets}
\label{app:NumberOfExecutionTickets}

To ballpark the "sufficiently" large number of tickets defined in \citep{burian_execution_2024} we run the following thought exercise. \citep{burian_execution_2024} states in Theorem 4: "\textit{If $n$ is sufficiently large, the current market cap of all tickets equals the present value of all future EL Rewards}".

The proof of the theorem from \citep{burian_execution_2024} follows the following steps:
\[ \lim _{n \rightarrow \infty} E\left[V_{\text {Issued Tickets }}\right]=\lim _{n \rightarrow \infty} n E\left[V_{\text {Ticket }}\right]=\lim _{n \rightarrow \infty}\left(\frac{n \mu_{R}}{(n d+1)}\right)=\lim _{n \rightarrow \infty}\left(\frac{\mu_{R}}{\left(d+\frac{1}{n}\right)}\right)=\frac{\mu_{R}}{d}=N P V_{R} \]

Hereby, we have the following:

\(\begin{array}{ll}
n &= \text{number of tickets} \\
\mu_{R} &= \text{expected value of } R \\
R &= \text{Random variable describing the prize for victory at time } t \\
d &= \text{inter-slot discount rate}
\end{array}\)

The theorem leads to the question of the definition of "\textit{sufficiently large}". To arrive at a ballpark number we propose the following. We introduce the following assumptions:

For the expected rewards $\mu_{R}$ we historically observe that the MEV-Boost payment per slot has been stable around 0.05 ETH with a high variance \citep{wahrstatter_time_2023}. At time of this writing this equals to around $\$ 175$\footnote{As of 18/03/2024}.

Assuming 2,628,000 block slots per year and using the current federal reserve interest rate of $5.5\%$\footnote{\url{https://de.wikipedia.org/wiki/Federal_Funds_Rate} (retrieved 18/03/2024)} we can ballpark $d$ :
$$
\begin{aligned}
1+d_{annual} &= \left(1+d_{Slot}\right)^{\textit{Slots per Year}} \\
d_{Slot} &= \left(1+d_{Annual}\right)^{1 / \textit{Slots per Year}}-1 \\
d_{Slot} &= 2.03732 e^{-8}
\end{aligned}
$$

Assuming that a deviation of $p_{var}$ from the present value of all future EL rewards is acceptable we can arrive at:
$$
\begin{aligned}
& N P V_{R}-E\left[V_{\text {Issued Tickets }}\right] \leq p_{\text {Var }} * N P V_{R} \\
& \frac{\mu_{R}}{d}-n * \frac{\mu_{R}}{\left(n^{*} d+1\right)} \leq p_{\text {Var }} * \frac{\mu_{R}}{d} \\
& \frac{\mu_{R}}{d}-\frac{\mu_{R}}{d+\frac{1}{n}} \leq p_{\text {Var }} * \frac{\mu_{R}}{d}
\end{aligned}
$$

This leads to:
\[
\begin{aligned}
& \frac{1}{\left(\frac{d}{1-p}-d\right)} \leq n \\
& \frac{1-p}{d^{*} p} \leq n
\end{aligned}
\]

To ballpark we can now insert the following values:
$$
\begin{aligned}
    d &= 2.03732 e^{-8} \\
    p_{var, 0.05} &= 0.05 \\
    p_{var, 0.5} &= 0.5
\end{aligned}
$$

This leads to:
$$
\begin{aligned}
& n_{0.05} \approx 932597726 \\
& n_{0.5} \approx 49084091 \\
& n_{0.9} \approx 5453787
\end{aligned}
$$

Divided by $2,628,000$ slots per year to: ca. 355 years (18 years for $n_{0.5}$ and 2.1 for $n_{0.9}$ respectively).

This leads to the preliminary conclusion that the number of tickets needs to be large, around 1 billion tickets to reliably capture $95\%$ of the net present value. This would be the amount of tickets covering roughly 350 years of slots. To capture $50 \%$ of the total value of execution layer rewards roughly 50 million tickets are needed, covering 18 years. To have the current market capitalization of issued tickets capture $10 \%$ of the net present value of all future execution layer rewards, around 5 million tickets covering 2 years are needed.

Note that this is extremely sensitive to the discount rate assumption, e.g. if a yearly discount rate of $20 \%$ is assumed the number of tickets decreases to $273,867,952$ (104 years) to capture $95 \%$ of all future execution layer rewards (for $p_{var, 0.5}: 14,414,103$ tickets, 5.5 years).

In designing the mechanism this heuristic can be kept in mind as a guidance to decide how much of all future execution layer rewards shall be captured in the current market capitalization. Capturing a high amount of all future execution layer rewards in the current market cap economically depends on the question whether protocol token holders or Execution Ticket holders have a higher discount rate. As it is to assume that discount rates differ for market participants and might be hard to reliably estimate, answering this question might be difficult. Hence, it might make sense to remain neutral on this.

Another dimension to consider is the cost of centralization. As \citep{burian_execution_2024} outlined, a too low number of tickets might easily capture a high percentage of tickets.

Given the formula outlined in Theorem 3 by \citep{burian_execution_2024} stated as
\[ E\left[V_{\text {all tickets }}\right]=N P V_{R}=\frac{\mu_{R}}{d} \]

and based on the assumptions above regarding $d$ and setting $\mu_{R}$ at the historical value of $\$ 175$ we arrive at a net present value for all future execution layer MEV rewards of
\[ NPV_{R} \approx 2,674,349 \:\Xi\: (\textit{at current rate } \$8,589,715,901) \]

Given the low assumption regarding the discount factor this can be seen as an upper limit. This can be used as a baseline for considering what the costs of centralization should be. If the cost to control $51 \%$ of issued tickets should be at least $1 \%$ of all future net present value we arrive at a target of roughly 1 million tickets with a market capitalization of all outstanding tickets of around $\$ 175$ million. Another more cosmetic argument for around one million tickets is the fact that this is also roughly the current number of validators as mentioned by \citep{drake_session_2023}. So, the chance of getting selected as a validator running one node currently or as a future Execution Ticket holder with one ticket, would be roughly equal.

Furthermore, the number of issued tickets interplays with the lookahead period, as: 
\[ \% P[Allocated~Tickets] = \frac{Slots~in~lookahead}{n}\] 
With a high number of $\%P[Allocated~Tickets]$ the risk for multi-MEV might be higher, as the certainty of slot allocation is higher.